\documentclass[12pt,preprint]{aastex}

\shorttitle{Fast Solar Wind}
\shortauthors{van Ballegooijen \& Asgari-Targhi}

\begin{document}

\title{Heating and Acceleration of the Fast Solar Wind\\
by Alfv\'{e}n Wave Turbulence}

\author{A. A. van Ballegooijen\altaffilmark{1},
M. Asgari-Targhi\altaffilmark{1} }
\altaffiltext{1}{Harvard-Smithsonian Center for Astrophysics, 
60 Garden Street, Cambridge, MA 02138, USA}

\begin{abstract}
We present numerical simulations of reduced magnetohydrodynamic (RMHD) turbulence
in a magnetic flux tube at the center of a polar coronal hole. The model for the
background atmosphere is a solution of the momentum equation, and includes the
effects of wave pressure on the solar wind outflow. Alfv\'{e}n waves
are launched at the coronal base, and reflect at various heights due to variations
in Alfv\'{e}n speed and outflow velocity. The turbulence is driven by nonlinear
interactions between the counter-propagating Alfv\'{e}n waves. Results are presented
for two models of the background atmosphere. In the first model the plasma density
and Alfv\'{e}n speed vary smoothly with height, resulting in minimal wave reflections
and low energy dissipation rates. We find that the dissipation rate is insufficient
to maintain the temperature of the background atmosphere. The standard phenomenological
formula for the dissipation rate significantly overestimates the rate derived from
our RMHD simulations, and a revised formula is proposed.
In the second model we introduce additional density variations along the flux tube with
a correlation length of 0.04 $R_\odot$ and with relative amplitude of $10 \%$. These
density variations simulate the effects of compressive MHD waves on the Alfv\'{e}n waves.
We find that such variations significantly enhance the wave reflection and thereby the
turbulent dissipation rates, producing enough heat to maintain the background atmosphere.
We conclude that interactions between Alfv\'{e}n- and compressive waves may play
an important role in the turbulent heating of the fast solar wind.
\end{abstract}


\clearpage

\section{Introduction}

The fast solar wind has velocities in the range 500 to 800 $\rm km ~ s^{-1}$,
and originates in coronal holes, which are low-density, open magnetic
field structures in the Sun's corona \citep[see][]{Zirker1977}.
The fast wind is believed to be driven by Alfv\'{e}n waves that are
launched in the photosphere and propagate outward along the open field
lines \citep[e.g.,][]{Parker1965, Heinemann1980, Velli1993, Dmitruk2002, 
Suzuki2005, Suzuki2006, Cranmer2007, Verdini2009, Cranmer2015}. 
Alternatively, the waves may be launched by micro-flares in the
chromospheric network \citep[e.g.,][]{Axford1992}.
Direct evidence for the existence of Alfv\'{e}n waves comes from {\it in situ}
observations of fluctuations in the solar wind \citep[e.g.,][]{Coleman1968,
Belcher1971, Bale2005}. The waves have a broad spectrum of wavenumbers and
frequencies, and are believed to be turbulent, that is, the different
wave modes interact nonlinearly with each other, and the wave energy
is continually being transferred from larger to smaller spatial scales
\citep[e.g.,][]{Hollweg1982, Hollweg1986, Matthaeus1990}. In the inertial
range the turbulence is highly anisotropic:
the fluctuating field $\delta {\bf B}$ is nearly perpendicular to the
background field ${\bf B}_0$, and varies more rapidly in the direction
perpendicular to ${\bf B}_0$ than along it. Alfv\'{e}n waves have also been
detected by studying wave-like phenomena in the solar atmosphere
\citep[][]{Tomczyk2007, Tomczyk2009, DePontieu2007, Morton2015}. These
observations provide evidence for the existence of counter-propagating
Alfv\'{e}n waves in the corona. The Interface Region Imaging Spectrograph
(IRIS) has observed signatures of Alfv\'{e}n waves carried by network jets
in the transition region of coronal holes \citep[][]{Tian2014}, indicating
that Alfv\'{e}n waves are injected into the corona at the base.

The mechanisms by which the fast solar wind is heated and accelerated
are not yet fully understood. Alfv\'{e}n waves are the most promising wave
type for transporting energy over large distances in the corona and solar
wind \citep[][]{Barnes1966, Velli1989, Matthaeus1999, Suzuki2006, Chandran2011}. 
However, it is not clear how the wave energy is transferred to the particles.
Some authors have argued that the waves have very high frequencies ($\sim$ kHz)
and interact with the particles via ion-cyclotron resonances \citep[][]{Tu1995}.
Such models can explain why the different ion species have different
temperatures and outflow velocities, but it is not clear how the
ion-cyclotron waves would be produced. Others have suggested that the
coupling between Alfv\'{e}n- and compressive waves plays an important role
in Alfv\'{e}n-wave dissipation \citep[][]{Kudoh1999, Moriyasu2004, Suzuki2005,
Suzuki2006, Matsumoto2010}.

Many authors have focused on wave turbulence as a mechanism for coronal
heating \citep[e.g.,][]{Hollweg1986, Velli1989, Matthaeus1999}.
One likely source of turbulence is the nonlinear interaction between
counter-propagating Alfv\'{e}n waves \citep[][]{Iroshnikov1963, Kraichnan1965}.
The waves can be described in terms of Elsasser variables, ${\bf z}_{\pm}
\equiv {\bf v}_1 \mp {\bf B}_1 / \sqrt{4 \pi \rho_0}$, where ${\bf B}_1$ and
${\bf v}_1$ are the magnetic and velocity fluctuations of the waves, and
$\rho_0$ is the mean plasma density.
In a homogeneous medium the ${\bf z}_{+}$ and ${\bf z}_{-}$ fields would
represent outward and inward-propagating waves, respectively. However, in an
inhomogeneous atmosphere such as the corona the Alfv\'{e}n speed $v_A (r)$
varies with radial distance $r$, and the ${\bf z}_{+}$ and ${\bf z}_{-}$ waves
are linearly coupled to each other \citep[][] {Heinemann1980, Velli1993,
Hollweg2007}. Therefore, the ${\bf z}_{+}$ waves launched by the Sun
naturally produce ${\bf z}_{-}$ waves at larger heights. \citet[][]{Velli1989}
have shown that in an inhomogeneous atmosphere with dominantly outward waves
the ${\bf z}_{+}$ field generates a ``secondary component" in the ${\bf z}_{-}$
field, which travels in the same direction as the ${\bf z}_{+}$ primary
field. Therefore, in general the ${\bf z}_{-}$ field has both inward and
outward-propagating components \citep[also see section 5 in][]{Perez2013}.
In this paper we will refer to the ${\bf z}_{-}$
waves as the ``minority" wave type, as their amplitudes are generally much
smaller than those of the dominant ${\bf z}_{+}$ waves. The nonlinear
interactions between dominant and minority waves create turbulence and cause
heating of the coronal plasma \citep[][]{Matthaeus1999}.

Phenomenological models for Alfv\'{e}n-wave turbulence have been developed in
which the plasma heating rate $Q_A$ is determined by the root-mean-square (rms)
values of the Elsasser variables, $Z_{\pm} \equiv \sqrt{< | {\bf z}_\pm |^2 >}$,
where $< \cdots >$ represents a spatial average \citep[e.g.,][]{Zhou1990,
Hossain1995, Matthaeus1999, Dmitruk2001, Dmitruk2002}. \citet[][]{Cranmer2005}
and \citet[][]{Cranmer2007} used such a phenomenological model to develop
a comprehensive description of the Alfv\'{e}n waves in the solar atmosphere and
fast wind, and compared the wind model with observations \citep[also see][]
{Verdini2007, Verdini2010}. \citet[][]{Chandran2011}
further improved the wind model by including separate energy equations for the
electrons and protons, and the proton temperature anisotropy. A more advanced
approach to turbulence modeling is the so-called shell model
\citep[][]{Buchlin2007, Verdini2009, Verdini2012} in which the wave spectrum
is described in more detail but the nonlinear terms are still approximated.

An even more complete description of turbulence is provided by direct numerical
simulations using the reduced magnetohydrodynamic (RMHD) equations.
\citet[][]{Oughton2001} and \citet[][]{Dmitruk2003} performed RMHD simulations
for open magnetic fields, and showed that reflection-driven turbulence can be
maintained in such structures. These authors neglected the effects of the solar
wind outflow on the waves. Similarly, \citet[][hereafter paper~I]{vanB2011}
simulated Alfv\'{e}n wave turbulence in coronal loops, but neglected the effects
of flows along the loops \citep[also see] [hereafter papers~II and III]
{Asgari2012, Asgari2013}. \citet[][]{Perez2013} were the first to perform RMHD
simulations for the fast solar wind, including the effects of outflow.
They solved the RMHD equations in a simulation domain that consists of a narrow
magnetic flux tube with a square cross section extending from the coronal base
($r = 1$ $R_\odot$) out to the Alfv\'{e}n critical point, which in their model
is located at $r_{\rm A} = 11.1$ $R_\odot$. 
\citet[][]{Perez2013} considered two values for the domain size
$L_{\perp \odot}$ at the coronal base (10 Mm and 20 Mm), and different values for
the correlation time of the injected waves (between 2 and 22 min).
The waves are launched with dimensionless perpendicular wavenumbers in the range
$1 \le \tilde{k}_\perp \le 3$, which corresponds to actual wavenumbers $k_\perp =
2 \pi \tilde{k}_\perp / L_{\perp \odot}$.
Up to one third of the wave energy launched at the base is dissipated in the corona
below the Alfv\'{e}n critical point, and another third goes into doing work on the
solar wind outflow. The remainder escapes into the heliosphere beyond the
Alfv\'{e}n point.

In their RMHD simulations, \citet[][]{Perez2013} found peak heating rates
ranging from $10^{10}$ $\rm erg ~ g^{-1} s^{-1}$ to $3 \times 10^{10}$
$\rm erg ~ g^{-1} s^{-1}$. This is somewhat lower than the values found in
earlier wave-driven wind models \citep[][]{Cranmer2007, Verdini2010,
Chandran2011}, where the peak heating rates are about $3 \times 10^{11}$
$\rm erg ~ g^{-1} s^{-1}$. The latter represents the heating rate needed to
raise the coronal temperature to the observed value of about 1 MK, despite
the strong cooling associated with thermal conduction and the solar wind
expansion. Given that RMHD simulations provide a more accurate description of
the turbulence, this suggests that the earlier models may have overestimated
the ability of Alfv\'{e}n wave turbulence to provide the required heating.

The purpose of the present paper is to further test the hypothesis that
reflection-driven wave turbulence can provide the energy needed for heating the
coronal plasma in the acceleration region of the fast solar wind. Following
\citet[][]{Perez2013}, we use RMHD simulations to describe the wave turbulence.
The basic model is described in Section 2, but much of the details (including all
of the equations) are presented in four Appendices. Simulation results are presented
in Section 3, where the simulated wave dissipation rates are compared with those
needed to sustain the background atmosphere. In Section 4 we consider the effects
of density fluctuations, which may significantly increase the turbulent heating
rate \citep[][]{Raymond2014}. In Section 5 we present a phenomenological model
for the turbulent dissipation rate. In Section 6 we discuss the
implications of our results for understanding the heating and acceleration of
the fast solar wind.

\section{Model for Wave-Driven Wind in an Open Flux Tube}
\label{sect:MHD}

In this paper we consider a thin magnetic flux tube extending along the solar
rotation axis at the center of the North polar coronal hole around the time of
cycle minimum. We first describe the global magnetic field in which the flux tube
is embedded. For this purpose we use a spherical coordinate system $(r,\theta,\phi)$,
where $r$ is the radial distance from Sun center and $\theta$ is the polar angle.
The magnetic field ${\bf B}$ is assumed to be axisymmetric with magnetic vectors
that lie in meridional planes ($B_\phi = 0$). We assume that a current sheet is
located in the equatorial plane, consistent with observations of narrow coronal
streamers \citep[e.g.,][]{Wang1997}. For simplicity we assume that the current
sheet extends down to the coronal base, i.e., we ignore the fact that coronal
streamers have closed magnetic fields at lower height \citep[e.g.,][]{Riley2011}.
This approximation is adequate for our present purpose because the closed fields
in equatorial streamers at cycle minimum have only a minor effect on the magnetic
fields over the solar poles.
Above and below the equatorial plane the magnetic field is assumed to be potential,
${\bf B} = - \nabla \Phi$ with $\nabla^2 \Phi = 0$. For $r \gg R_\odot$ the
field becomes nearly radial and falls off like a monopole, $B_r \propto r^{-2}$.
The magnetic flux on the photosphere is assumed to be highly concentrated in the
polar regions. In the northern hemisphere $B_r (R_\odot,\theta) = B_{\rm pole}
\cos^8 \theta$, where $B_{\rm pole}$ is the net flux density at the pole
(we use $B_{\rm pole} = 10$ G). This flux distribution is
consistent with observations of the Sun's magnetic field near cycle minimum
\citep[][]{DeVore1984, Sheeley1989}.
Expanding $\cos^8 \theta$ in terms of Legendre polynomials $P_{2n-2}
(\cos \theta)$, we can extrapolate the field to larger heights. For the field
line along the rotation axis ($\theta = 0$) we find
\begin{equation}
B_0 (r) = \sum_{n=1}^{5} B_n \left( \frac{r}{R_\odot} \right)^{-2n} ,
\label{eq:B0poten}
\end{equation}
where $B_n = B_{\rm pole} [715, 2600, 2160, 832, 128 ] / 6435$ with
$n = 1, \cdots, 5$. The field strength $B_0 (r)$ inside the flux tube is
assumed to be equal to that of the background field.

The modeled flux tube extends along the solar rotation axis from the coronal
base outward into the heliosphere. The base is assumed to be located at
radial distance $r_{\rm base} = 1.003$ $R_\odot$, and we follow the tube out
to $r_{\rm max} = 20$ $R_\odot$. The tube is assumed to have a circular
cross-section with radius $R(r)$, and magnetic flux is conserved, so $R^2 B_0$
is constant along the tube. The radius of the flux tube at the coronal base is
assumed to be $R_{\rm base} = 1$ Mm, which implies that the tube is everywhere
thin compared to the solar radius. The flux tube is treated as having a rigid
outer boundary, and plasma flows do not penetrate this boundary. Unlike in our
previous work on coronal loops (papers I, II and III), the Alfv\'{e}n waves
are launched by imposing transverse motions on the plasma at the coronal base,
so the lower atmosphere is not included in the present model. The imposed
``footpoint" motions are assumed to be incompressible and confined to the
circular cross-section of the tube. Also, the footpoint motions are assumed
to have a velocity amplitude $v_{\rm rms} \approx 40$ $\rm km ~ s^{-1}$,
consistent with observed spectral line widths and non-thermal velocities in
coronal holes \citep[][]{Wilhelm1998, McIntosh2008, Banerjee2009, Landi2009,
Singh2011, Hahn2012, Bemporad2012}. The waves launched by these footpoint motions
travel upward within the modeled flux tube, and dissipate their energy over a wide
range of heights, heating the coronal plasma to temperatures $T_0 (r) \sim 1$ MK.
The resulting gas pressure gradients are an important factor in driving the
solar wind \citep[][]{Parker1958, Parker1960}. The wind has a mean outflow
velocity $u_0 (r)$
and mass density $\rho_0 (r)$. The Alfv\'{e}n waves also exert a direct force
on the plasma through the wave pressure gradient, which plays a key role in
producing the high speed of the wind emanating from corona holes 
\citep[][]{Belcher1971a, Alazraki1971, Hollweg1973, Jacques1977, Jacques1978,
Marsch1997}. Similar effects occur in other stars \citep[e.g.,][]{Hartmann1980}.
The goal of the present paper is to construct a model for the interactions
between the Alfv\'{e}n waves and the plasma in the fast solar wind.

The reasons for choosing $R_{\rm base} = 1$ Mm are as follows. Although the
lower atmosphere is not explicitly included in our modeling, we must take into
account that the magnetic field in the lower atmosphere is highly fragmented
and consists of discrete flux elements surrounded by more nearly field-free
plasma. In the photosphere these ``flux tubes" have kilogauss field strengths
and widths of order 100 km, and they are located in the intergranulation lanes
of the solar granulation pattern \citep[e.g.,][]{Stenflo1973, Title1987,
deWijn2009}. In the photospheric regions below a polar coronal hole the average
magnetic flux density is about 10 G, so the kilogauss flux tubes cover only
$1 \%$ of the available area. The flux tubes expand with height in the solar
atmosphere, and neighboring flux tube ``merge" to form a more continuous field
in the low corona \citep[e.g.,][]{Cranmer2005}. A typical flux tube increases
in width from about 100 km in the photosphere to about 1000 km at the coronal
base (at height $\sim 2000$ km), hence our choice for the flux tube radius
$R_{\rm base}$. Also, the magnetic field strength drops from about 1000 G in the
photosphere to about 10 G in the low corona, which equals our value for the
field strength $B_{\rm pole}$ at the coronal base.

In the photosphere the magnetic flux tubes are continually shuffled about and
deformed by convective flows associated with the solar granulation
\citep[][]{Muller1994, Berger1996, Berger1998, Chitta2012}. These motions cause
magnetic disturbances inside the flux tubes that propagate upward in the form
of Alfv\'{e}n waves and/or kink waves \citep[e.g.,][]{Spruit1982, Edwin1983,
Morton2013}. Due to the density stratification of the lower atmosphere,
the waves are significantly amplified
on their way to the corona. The wave amplitudes increase from about
1 $\rm km ~ s^{-1}$ in the photosphere to about 40 $\rm km ~ s^{-1}$ in
the low corona, which is crucial for producing the wave amplitudes needed
to accelerate the solar wind \citep[][]{Cranmer2007}. Therefore, we believe
that the solar granulation is the main driver of the ``footpoint" motions
of the field lines in the low corona. These motions are not expected
to be coherent from one photospheric flux tube to another.
Therefore, the velocity auto-correlation length $\lambda_{\perp \odot}$
of the ``footpoint" motions must be less than the flux tube radius,
$\lambda_{\perp \odot} < R_{\rm base} \sim 1$ Mm. In this paper we estimate
$\lambda_{\perp \odot}$ as the inverse of the perpendicular wavenumber of
the imposed footpoint motions, $\lambda_{\perp \odot} = R_{\rm base} /
\tilde{k}_\perp$, where $\tilde{k}_\perp$ is the dimensionless perpendicular
wavenumber. For the flux tube models considered in this paper
$\tilde{k}_\perp = 3.832$, the first zero of the $J_1 (x)$ Bessel function,
which yields $\lambda_{\perp \odot} \approx 261$ km.
While the Sun has supergranular flows on scales of 10 to 30 Mm,
we believe that the correlation times of such flows (hours) are too long for the
associated waves to be amplified in the lower atmosphere. Therefore, we do not
believe that the supergranular flows can play a significant role in driving the
Alfv\'{e}n waves that heat and accelerate the fast solar wind.

We also require that the footpoint motions resemble a random walk, not a
persistent rotational motion. The dynamical time can be defined as the time
it takes for a footpoint to travel a distance equal to the flux tube diameter,
$\tau_{\rm dyn} = 2R_{\rm base} / v_{\rm rms} = 50$ s. In the present model
the footpoint motions are confined to the circular cross-section of the tube.
Therefore, if the correlation time $\tau_{\rm c}$ of the velocity is much
larger than $\tau_{\rm dyn}$, the footpoints move in a rotational pattern that
persists for many turns before the pattern changes. We believe such persistent
rotational motions are not realistic, given the randomness of the granule-scale
convective flows that are ultimately responsible for the ``footpoint" motions
at the coronal base.
Therefore, in this paper we assume $\tau_{\rm c} \approx 50$ s, comparable to
the dynamical time at the coronal base. This is significantly shorter than the
correlation times of 2 to 22 minutes assumed by \citet[][]{Perez2013}. It is
well known that short correlation times make the reflection less efficient
\citep[e.g.,][]{Dmitruk2003}, but in our view the value of $\tau_{\rm c}$
should be chosen on the basis of a model for the structure and dynamics of
magnetic elements in the lower atmosphere, not the efficiency of wave reflection
in the corona.

The main dissipation mechanism for the Alfv\'{e}n waves is assumed to be wave
turbulence. As the ${\bf z}_{+}$ waves propagate upward along the flux tube
they encounter spatial variations in Alfv\'{e}n speed $v_A (r)$, which causes
the generation of ${\bf z}_{-}$ waves. This coupling happens not only in the
chromosphere and transition region (where $v_A$ increases by two orders of
magnitude), but also in the corona where smaller changes in Alfv\'{e}n speed
occur. In general the ${\bf z}_{-}$ waves have both inward and
outward-propagating components \citep[][]{Velli1989, Perez2013}.
The ${\bf z}_{-}$ waves interact with the ${\bf z}_{+}$ waves via a well-known
nonlinear process \citep[][]{Shebalin1983, Goldreich1995, Goldreich1997,
Bhattacharjee2001, Maron2001, Cho2002}. These nonlinear
interactions create turbulence and cause a rapid transfer of wave energy
to smaller and smaller spatial scales in the direction transverse to the
background field. 
Eventually the waves reach such small scales that wave-particle
interactions become important, and the wave energy is converted into heat
of the background plasma. At present the details of these (collisionless)
dissipation processes are not well understood, but this is not necessary for
estimating the dissipation rate because in a turbulent plasma the energy
cascade rate is mostly determined by the dynamics of the plasma on large
spatial scales.
Therefore, for simulating the overall dynamics of the solar wind it is necessary
to include the effects of wave turbulence, but it is sufficient to describe
the waves with relatively low spatial resolution (i.e., the turbulent waves
are only partially resolved). The plasma is treated as a single fluid, i.e.,
differences in temperature or velocity between the various ion species are
neglected, and the electron temperature is assumed to be equal to the ion
temperature.

The Alfv\'{e}n waves are described in terms of their effect on the magnetic field
${\bf B} ({\bf r},t)$ and plasma velocity ${\bf v} ({\bf r},t)$, which are
functions of position ${\bf r}$ within the flux tube and time $t$. The plasma
motions are governed by the MHD equations:
\begin{eqnarray}
 & & \frac{\partial \rho} {\partial t} + \nabla \cdot ( \rho {\bf v} ) = 0 ,
\label{eq:mass} \\
 & & \rho \frac{d {\bf v}} {dt} = - \nabla p  + \frac{1}{4 \pi}
( \nabla \times {\bf B} ) \times {\bf B} - \rho \frac{G M_\odot}{r^2}
\hat{\bf r} + {\bf D}_v ,  \label{eq:dvdt} \\
 & & \rho \left[ \frac{d}{dt} \left( \frac{1}{\gamma -1} \frac{p}{\rho} \right) 
+ p \frac{d}{dt} \left( \frac{1}{\rho} \right) \right] =
Q_{\rm A} - Q_{\rm rad} - Q_{\rm cond} ,  \label{eq:heat} \\
 & & \frac{\partial {\bf B}} {\partial t} = \nabla \times
( {\bf v} \times {\bf B} ) + {\bf D}_m ,  \label{eq:dBdt}
\end{eqnarray}
where $d/dt$ is the co-moving time derivative, $\rho ({\bf r},t)$ is the mass
density, $p({\bf r},t)$ is the plasma pressure, $G$ the gravitational constant,
$M_\odot$ the solar mass, $\hat{\bf r}$ is the unit vector in the radial
direction, and ${\bf D}_v$ and ${\bf D}_m$ are dissipative terms. In the heat
equation (\ref{eq:heat}), $\gamma$ is the ratio of specific heat coefficients,
$Q_{\rm A} ({\bf r},t)$ is the plasma heating rate per unit volume,
$Q_{\rm rad} ({\bf r},t)$ is the radiative loss rate, and
$Q_{\rm cond} ({\bf r},t)$ is the conductive loss rate. The latter is given
by the divergence of the conductive flux, $Q_{\rm cond} = \nabla \cdot
{\bf F}_{\rm cond}$, which may be positive or negative depending on position
in the corona. The conductivity tensor is highly anisotropic, so the conductive
flux ${\bf F}_{\rm cond}$ is nearly parallel to the magnetic field ${\bf B}$.
Since we only consider the coronal part of the flux tube, we neglect the
effects of partial ionization on the internal energy of the plasma,
$\gamma = 5/3$.

In this paper we use the Reduced MHD (RMHD) equations \citep[][]{Strauss1976,
Strauss1997} to simulate the dynamics of Alfv\'{e}n waves in the acceleration
region of the fast solar wind. The derivation of the RMHD equations for a thin
flux tube with variable field strength $B_0 (r)$ and density $\rho_0 (r)$ was
discussed in paper I. In Appendix A we discuss the modifications of the equations
needed to include the effects of the solar wind outflow $u_0 (r)$ on the waves
\citep[also see][]{Perez2013}. In Appendix B we describe the numerical methods
for solving these equations. We use a spectral method to describe the spatial
variations of the waves in the direction perpendicular to the mean magnetic
field, and finite-differences in the radial direction. The radial grid is chosen
such that for outward propagating waves the wave travel time between neighboring
grid points is constant with height and equal to the time step of the simulation,
$\Delta t = 1$ s. This allows the waves to propagate to large height without
numerically induced distortions. The key feature of the RMHD equations is that
they retain the nonlinear terms responsible for the development of Alfv\'{e}n
wave turbulence. To dissipate the waves, artificial damping is used at high
parallel and perpendicular wavenumbers.
For simplicity the same damping rates are used for both the magnetic- and
velocity fluctuations [see equations (\ref{eq:dhdt1}) and (\ref{eq:dodt1})],
so the magnetic Prandtl number (ratio of viscosity to magnetic diffusivity)
satisfies $Pr_{\rm M} = 1$.
In this paper we assume that all heating occurs via Alfv\'{e}n wave turbulence,
but it remains to be seen whether such heating is indeed sufficient to explain
the observed properties of the fast solar wind.

Unlike in our modeling of coronal loops in active regions (papers I, II and III),
we do not include the lower solar atmosphere in the present version of the
RMHD model. The main reason is that we need to simulate the waves for longer
periods than in our earlier loop simulations, and including the lower atmosphere
would require short time steps ($\Delta t \sim 0.1$ s), which is not practical
for simulating the solar wind with the present version of our RMHD code.
Therefore, the lower boundary of the RMHD model is assumed to be located at the
coronal base where the temperature is $T_0 \approx 0.3$ MK.

The outward- and inward propagating Alfv\'{e}n waves travel with velocities
$u_0 \pm v_A$, respectively. Hence, the waves are significantly affected by the
outflow when $u_0$ becomes comparable to $v_A$. In fact, at heights above the
Alfv\'{e}n critical point (where $u_0 > v_A$) the ``inward" propagating waves
actually move radially outward with a velocity less than the plasma velocity
$u_0$. Conversely, the Alfv\'{e}n waves have significant effects on the plasma:
(1) the wave turbulence causes plasma heating, and (2) the waves exert a direct
force on the out-flowing plasma via wave pressure gradients. The latter is thought
to play an important role in producing the high speed of the wind emanating from
coronal holes \citep[][]{Hollweg1973, Marsch1997}. Therefore, in the acceleration
region of the wind the plasma and waves must exchange energy with each other.
The details of this energy exchange are discussed in Appendix C.

Three-dimensional RMHD modeling requires that we first set up a one-dimensional
model for the background atmosphere inside the flux tube.
In addition to the magnetic field strength $B_0 (r)$ given in equation
(\ref{eq:B0poten}), we also need the plasma temperature $T_0 (r)$, density
$\rho_0 (r)$ and outflow velocity $u_0 (r)$. Appendix D describes how these
quantities are computed. The temperature is a prescribed function of $r$ and
is given by equation (\ref{eq:T0}). Using a formula for temperature has the
advantage that higher derivatives can be accurately computed, which is
important for evaluating the energy loss rates due to thermal conduction.
We require that the background model satisfy not only the mass conservation
equation (\ref{eq:mass0}) but also the equation of motion (\ref{eq:dudt0}).
The latter includes the wave pressure force $D_{\rm wp}(r)$, so we need an
approximation of this force that can be used in constructing the background
model. Appendix D describes how this approximation is obtained from the wave
action equation (\ref{eq:dSA}), which is an approximation for the wave energy
equation. We also describe how the approximation for $D_{\rm wp}$ is used in
the iterative process for deriving the outflow velocity $u_0 (r)$.

In this paper we distinguish between the wave energy dissipation rate
$Q_{\rm tot}$ and the plasma heating rate $Q_{\rm A}$. If the waves provide
all the heating, these two rates should be equal, $Q_{\rm tot} = Q_{\rm A}$.
The goal of the present work is to construct a solar wind model for which
this condition is indeed satisfied. If the condition is satisfied, we can
derive the full energy equation by adding equations (\ref{eq:E1}) and
(\ref{eq:E2}) from Appendix C:
\begin{equation}
B_0 \frac{d} {dr} \left( \frac{F_{\rm total}} {B_0} \right) = - Q_{\rm rad} ,
\label{eq:E3}
\end{equation}
where $F_{\rm total} = F_{\rm plasma} + F_{\rm waves}$ is the total energy
flux along the tube. It follows that in the absence of radiative losses the
total energy flow $\pi R^2 F_{\rm total}$ is constant along the tube.
This demonstrates the consistency of the equations used in this paper:
except for radiative losses, the total energy of the system (plasma + waves)
is conserved. However, in the next Section we will consider a model with
a smooth background atmosphere for which the condition $Q_{\rm A} =
Q_{\rm tot}$ is not satisfied. In this case the model is not consistent
from an energy point of view.

\section{Model with a Smooth Background Atmosphere}

\begin{figure}
\epsscale{1.0}
\plotone{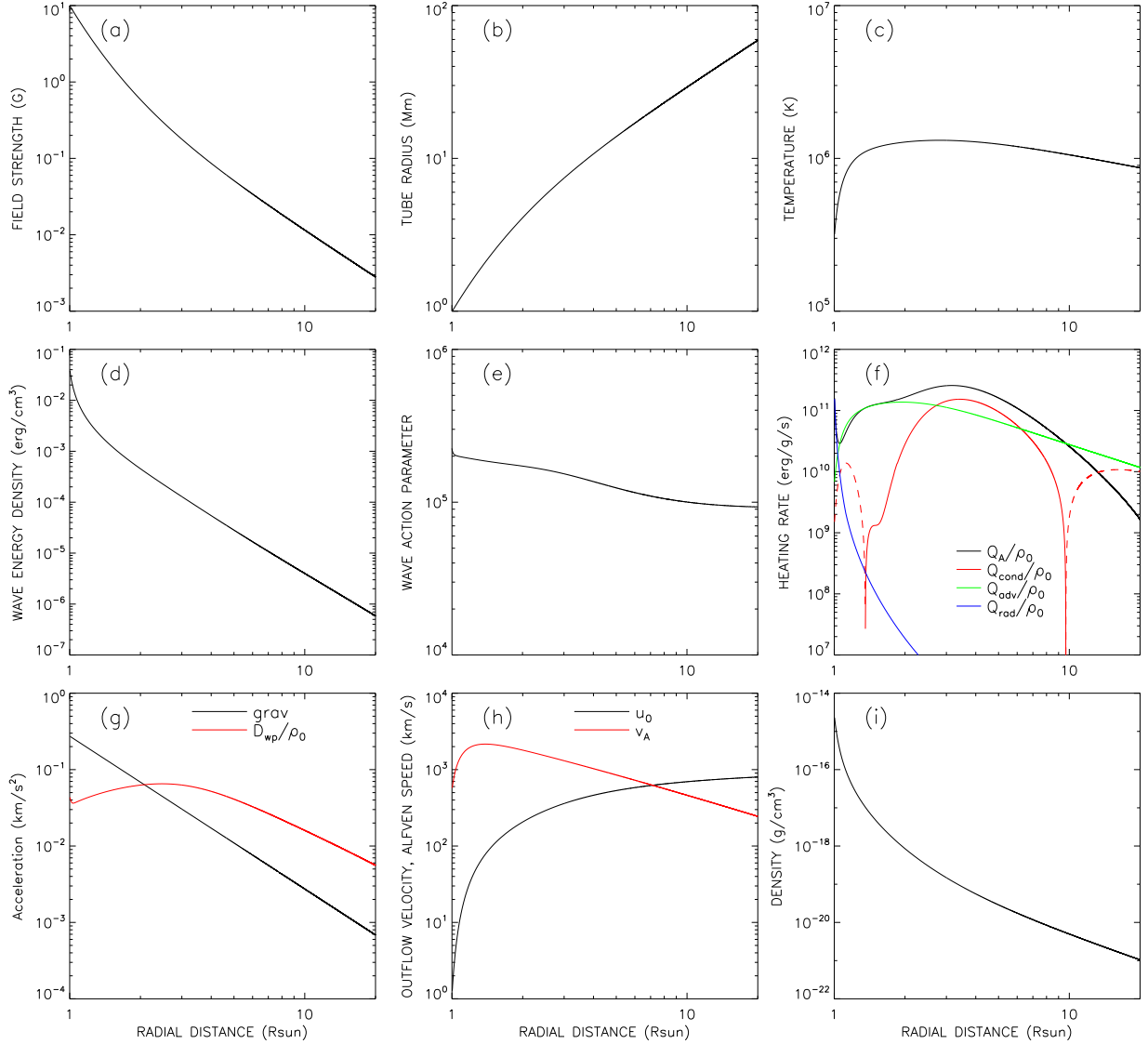}
\caption{Radial dependence of various background quantities for a polar
coronal hole.
(a) Magnetic field strength. (b) Flux tube radius. (c) Temperature.
(d) Wave energy density. (e) Wave action parameter. (f) Plasma heating rate
due to wave dissipation (black curve), and energy loss rates due to thermal
conduction (red curve), advection (green curve), and radiation (blue curve).
(g) Outward acceleration due to wave pressure gradient (red curve), and
inward acceleration due to gravity (black curve). (h) Outflow velocity
(black curve) and Alfv\'{e}n speed (red curve). (i) Mass density. }
\label{fig1}
\end{figure}

In this section we describe results for a reflection-driven wave-turbulence model
of the fast solar wind. Figure \ref{fig1} shows the structure of the background
atmosphere. Figure \ref{fig1}(a) shows the magnetic field strength $B_0 (r)$ as
computed from equation (\ref{eq:B0poten}) with $B_{\rm pole} = 10$ G. Note that
$B_0 (r)$ drops off faster than $r^{-2}$, which is due to the $\cos^8 \theta$
distribution of magnetic flux on the photosphere. Figure \ref{fig1}(b) shows the
flux tube radius $R(r)$, which is computed from flux conservation. Note that
$R(r)$ increases from 1 Mm at the coronal base to 60 Mm at $r = 20$ $R_\odot$.
Figure \ref{fig1}(c) shows the temperature $T_0 (r)$ computed from equation
(\ref{eq:T0}) with the parameters $C_0 = 0.35$, $C_1 = 2$, $m = 0.3$, and
$k = 8$. This formula is intended to give only a qualitative description of the
temperature structure in coronal holes, and does not represent an atmosphere in
thermal balance. Note that $T_0 (r)$ increases from about 0.3 MK at the
coronal base, and reaches a maximum of about 1.31 MK at $r \approx 3$ $R_\odot$.
At large heights the temperature decreases as $T_0 \propto r^{-0.3}$. 

All other quantities shown in Figure \ref{fig1} are computed as described in
Appendix D. We include the effects of the wave pressure force $D_{\rm wp}$
on the background plasma. Figure \ref{fig1}(d) shows the wave energy density
$U_{\rm A} (r)$, which is determined by solving the wave action equation
(\ref{eq:dSA}). The assumed amplitude of the waves at the coronal base is
$v_{\rm rms} (r_{\rm base}) = 40.6$ $\rm km ~ s^{-1}$. Note that
$dU_{\rm A}/dr < 0$, which implies that the waves produce an outward force on
the plasma ($D_{\rm wp} > 0$). Figure \ref{fig1}(e) shows the wave action
parameter $S_{\rm A} (r)$ defined in equation (\ref{eq:SA}). The wave action
decreases by about $50 \%$ over the height range of the model, which is due to
wave dissipation. For the purpose of constructing the background model the wave
dissipation rate is assumed to be equal to the plasma heating rate $Q_{\rm A}$,
and is computed by evaluating the energy losses of the plasma [see equations
(\ref{eq:heat0}) through (\ref{eq:Qcond})].
Figure \ref{fig1}(f) shows $Q_{\rm A} (r)$ together with its contributions from
radiative losses (blue curve), thermal conduction (red curve), and the advection
terms in the heat equation (green curve). Note that radiative losses are important
only in the low corona, and advection dominates in the region 1.1 - 2.7 $R_\odot$.
At $r > 10$ $R_\odot$ the conductive losses are negative (dashed red curve),
indicating the plasma is heated by conduction from below. Figure \ref{fig1}(g)
shows the acceleration $D_{\rm wp} / \rho_0$ due to the wave pressure force
(red curve) together with the acceleration of gravity (black curve). Note that
the wave pressure acceleration exceeds the gravitational acceleration already at
$r = 2$ $R_\odot$.

Figure \ref{fig1}(h) shows the outflow velocity $u_0 (r)$ (black curve)
as computed with the iterative method described in Appendix D, and Figure
\ref{fig1}(i) shows the corresponding density $\rho_0 (r)$ from mass flux
conservation. Here we assumed a coronal base pressure $p_{0,\rm base} = 0.1$
$\rm dyne ~ cm^{-2}$, which yields a base density $\rho_{0,\rm base} =
2.32 \times 10^{-15}$ $\rm g ~ cm^{-3}$. The Alfv\'{e}n speed $v_A (r)$ is
plotted as the red curve in Figure \ref{fig1}(h), and reaches a maximum of
about 2000 $\rm km ~ s^{-1}$ at $r = 1.5$ $R_\odot$. The red and black curves
cross at $r = 7.13$ $R_\odot$, which is the Alfv\'{e}n critical point. At the
outer boundary of the model ($r = 20$ $R_\odot$) the outflow velocity reaches
about 800 $\rm km ~ s^{-1}$, typical for the terminal velocity of the fast
solar wind. The critical point of the flow is located at $r_{\rm c} = 1.83$
$R_\odot$, which is significantly smaller than the value $r_{\rm c} = 4.48$
$R_\odot$ that would exist in the absence of wave pressure forces (i.e.,
with only $dp_0/dr$ driving the flow). Therefore, the wave pressure force
$D_{\rm wp}$ plays a major role in producing the fast solar wind
\citep[e.g.][]{Belcher1971a, Hollweg1973, Jacques1978}.

\subsection{RMHD Simulations}

The atmospheric model presented in Figure \ref{fig1} is used as the
``background" for three-dimensional, time-dependent RMHD simulations of the
Alfv\'{e}n waves inside the flux tube. The numerical methods for solving the
RMHD equations are described in Appendix B. In the initial state there are
no waves at any height. The waves are launched by imposing certain
``footpoint" motions on the plasma at the coronal base, which is the lower
boundary of the RMHD model. The imposed velocity patterns are a combination
of two basis functions $F_k (\xi,\varphi)$ as described in Appendix B
(also see Appendix B of paper~I). In our model these ``driver modes" have
indices $k=10$ and $k=11$, and both modes have dimensionless wavenumber
$a_k = 3.832$, given by the first zero of the $J_1 (x)$ Bessel function.
Also, both modes have azimuthal mode number $m_k = 1$, but with different
directions of the flow in the perpendicular plane. The amplitudes
$f_k (r_{\rm base},t)$ of the driver modes vary randomly with time $t$ in the
simulation. For each mode we first create a normally distributed random
sequence $f(t)$ on a grid of times covering the entire simulation
($t_{\rm max} =$ 30,000 s). Then the sequence is Fourier filtered using
a Gaussian function $G( \tilde{\nu} ) = \exp [-(\tau_0 \tilde{\nu} )^2]$,
where $\tilde{\nu}$ is the temporal frequency (in Hz) and $\tau_0$ is a
specified parameter. In the present work we take $\tau_0 = 120$ s, which
corresponds to a correlation time $\tau_{\rm c} = \tau_0 / \sqrt{2 \pi}
\approx 48$ s. The filtered sequence is renormalized such that the rms
vorticity of each mode is $\omega_{\rm rms} = 0.11$ $\rm rad ~ s^{-1}$;
this corresponds to a combined velocity amplitude $v_{\rm rms} = 40.6$
$\rm km ~ s^{-1}$, consistent with the value used in the setup of the
background model. Note that the correlation time $\tau_{\rm c}$ is
comparable to the dynamical time of the footpoint motions, $\tau_{\rm dyn}
= 2R_{\rm base} / v_{\rm rms} = 49.3$ s. Therefore, the footpoints are
strongly intermixed on the correlation timescale, leading to strong
turbulence in the Alfv\'{e}nic fluctuations.

The simulated waves are described in terms of mode amplitudes $h_k (r,t)$
and $f_k (r,t)$, which represent the magnetic- and velocity fluctuations,
respectively (see Appendix B). The waves are also described with Elsasser-like
variables, $\omega_{\pm,k} (r,t)$, which represent the vorticities in the
${\bf z}_{\pm}$ waves. The dominant $\omega_{+,k}$ waves always travel outward,
but the minority $\omega_{-,k}$ waves can have both inward and outward-propagating
components \citep[][]{Velli1989}. The linear coupling between the $\omega_{+,k}$
and $\omega_{-,k}$ waves is described by the second and third terms in equation
(\ref{eq:omeg_pm}). This coupling is due to spatial variations in Alfv\'{e}n
speed $v_A (r)$, density $\rho_0 (r)$, field strength $B_0 (r)$, and outflow
velocity $u_0 (r)$.
The outward-propagating waves first reach the outer boundary of the RMHD model
($r = 20$ $R_\odot$) after about 10,891 s. The waves are simulated for a period
of 30,000 s to ensure that a statistically stationary state is reached.
In this state there are dominant $\omega_{+,k}$ waves and minority
$\omega_{-,k}$ waves at all heights, and the waves have a broad spectrum of
perpendicular wavenumbers, indicating that strong wave turbulence has developed.
We find that the minority waves mainly travel outward with the same velocity
($u_0 + v_A$) as the dominant waves, consistent with the predictions of
\citet[][]{Velli1989}. Therefore, it is not correct to think of the
$\omega_{-,k}$ waves as inward-propagating waves.

\begin{figure}
\epsscale{1.0}
\plotone{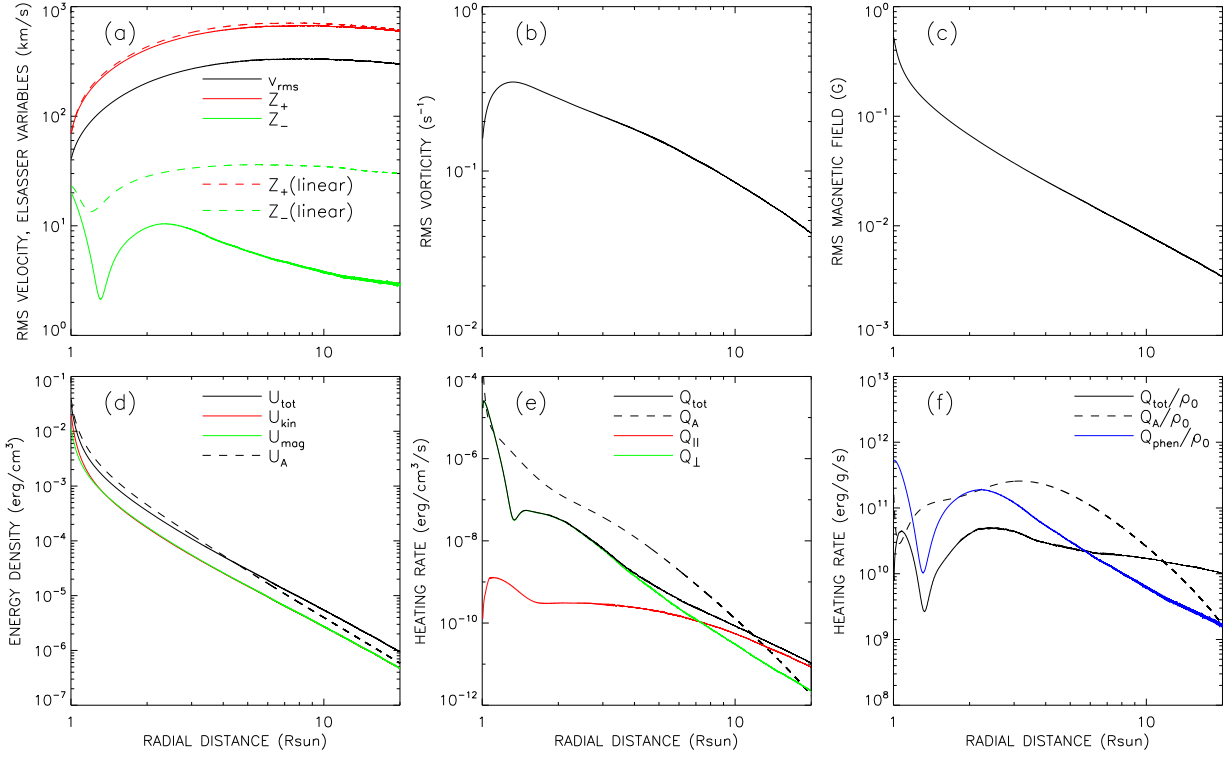}
\caption{Radial dependence of various wave-related quantities for a polar
coronal hole model, as derived from RMHD simulation. (a) Velocity amplitude
of the waves (black curve), and Elsasser variables for dominant waves
(red curve) and minority waves (green curve). The dashed red/green curves
are for a model with the nonlinear terms switched off. (b) Amplitude
of the parallel component of vorticity. (c) Amplitude of the fluctuating
component of magnetic field. (d) Wave energy densities: total energy (black
curve), kinetic energy (red curve), and magnetic energy (green curve).
Also shown is the wave energy density assumed in setup of the background
atmosphere (dashed curve). (e) Wave energy dissipation rates per unit volume:
total wave dissipation rate $Q_{\rm tot}$ (solid black curve), 
together with contributions from $Q_{\perp}$ (green curve) and
$Q_{\parallel}$ (red curve).
Also shown is the plasma heating rate $Q_{\rm A}$ assumed in setup of the
background atmosphere (dashed black curve).
(f) Wave energy dissipation rates per unit mass: rate derived from turbulence
simulation (solid black curve), rate assumed in the setup of background
atmosphere (dashed curve), and rate predicted by the phenomenological model
of equation (\ref{eq:Qphen3}) (blue curve).}
\label{fig2}
\end{figure}

Figure \ref{fig2} shows various wave-related quantities averaged over the
cross-section of the flux tube and over the time. Each quantity is averaged
over the time interval $t_0 (r)+300 \le t \le 30000$ (in seconds), where
$t_0 (r)$ is the time for an
outward propagating wave to reach a certain height. The black curve in Figure
\ref{fig2}(a) shows the rms velocity amplitude of the waves, $v_{\rm rms} (r)$,
which reaches a peak value of about 330 $\rm km ~ s^{-1}$ at $r \approx 9$
$R_\odot$. Comparing with Figure \ref{fig1}(h), we see that $v_{\rm rms} < v_A$
at most heights, but $v_{\rm rms} \sim v_A$ near the outer boundary of the 
model at $r = 20$ $R_\odot$. This indicates that the RMHD approximation begins
to break down at that height. The {\it solid} red and green curves in Figure
\ref{fig2}(a) show the rms values of the Elsasser variables, $Z_{\pm} (r) =
\sqrt{ < | {\bf z}_\pm |^2 >}$, where ${\bf z}_{\pm} \equiv {\bf v}_1 \mp
{\bf B}_1 / \sqrt{4 \pi \rho_0}$. Note that the minority $Z_{-}$ waves are
are much weaker than the dominant $Z_{+}$ waves; at $r > 5$ $R_\odot$ the ratio
$Z_{-} / Z_{+} < 0.01$, consistent with the results of \citet[][]{Cranmer2007}.
The amplitude $Z_{-} (r)$ of the minority waves has a sharp minimum at
$r \approx 1.3$ $R_\odot$. We attribute this to the fact that the Alfv\'{e}n
speed has a maximum near that height [see Figure \ref{fig1}(h)], which reduces
the magnitude of the second term in equation (\ref{eq:omeg_pm}) and thereby
the amplitude of the minority waves. 
The Elsasser variable of the minority waves, $Z_{-}$, is significantly
smaller in our model compared to \citet[][]{Perez2013}, probably because we
use driver waves with shorter correlation times (i.e., shorter wavelength and less
reflection).

The {\it dashed} red and green curves in Figure \ref{fig2}(a) give the Elsasser
variables for a different ``linear" model in which the nonlinear and damping
terms in the RMHD equations (\ref{eq:dhdt1}) and (\ref{eq:dodt1}) are omitted.
In that case there is no turbulent cascade, so the level of minority waves is
determined solely by the linear (reflection) terms in the equations. Comparing the
solid and dashed green curves in Figure \ref{fig2}(a), we see that one effect
of the nonlinear terms is to suppress the amplitude of the minority waves by
a factor of 3 to 10 compared to the linear model. This comparison between the
two models shows that the amplitude of the minority waves is determined by two
processes: (1) linear coupling with the dominant waves, and (2) decay of the
minority waves due to turbulence. The linear process (1) can lead to either
production or destruction of minority waves, depending on whether the ratio
$Z_{-}/Z_{+}$ is smaller or larger than the value $(Z_{-}/Z_{+})_{\rm lin}$
found in a purely linear model (i.e., model without nonlinear terms). In such
a linear model only the first process operates, and a certain level of minority
waves is obtained, as shown by the dashed green curve in Figure \ref{fig2}(a).
This level of $Z_{-}$ is already low compared to $Z_{+}$ because the reflection
is relatively weak in our ``smooth" model. With the nonlinear terms switched on,
$Z_{-}$ is further reduced as shown by the solid green curve. This additional
reduction is due to the second process, turbulent decay of minority waves.
The reduction is significant because the minority waves have a short nonlinear
time scale (see Section 5). Therefore, the nonlinear interactions play an
important role in determining the amplitudes of the minority waves
\citep[also see][]{Chandran2010}.

Figure \ref{fig2}(b) shows the rms vorticity of the waves, i.e., the component
of vorticity parallel to the background field. This quantity is dominated by waves
with higher perpendicular wavenumbers, and therefore is sensitive to the spatial
resolution of the model. Note that the vorticity decreases with height, which
is due to the expansion of the flux tube with height (a similar effect was found
for coronal loops, see paper II). Figure \ref{fig2}(c) shows the rms value of
the magnetic fluctuations of the waves, $B_{1,\rm rms} (r)$. Comparing with
Figure \ref{fig1}(a), we see that $B_{1,\rm rms} \approx B_0$ near the outer
boundary of the model, again indicating that the RMHD approximation begins to
break down at that height.

Figure \ref{fig2}(d) shows the total energy density $U_{\rm tot}$ of the simulated
waves (full black curve), together with the contributions to this quantity from
the kinetic energy $U_{\rm kin}$ (red curve) and magnetic energy density
$U_{\rm mag}$ (green curve). The dashed curve shows the energy density $U_{\rm A}$
used in the setup of the background model [same as Figure \ref{fig1}(d)]. We see
that $U_{\rm kin} \approx U_{\rm mag}$ and $U_{\rm tot} \approx U_{\rm A}$, so
the assumptions made in the model setup (see Appendix D) seem consistent with the
wave simulation results.

Figure \ref{fig2}(e) shows the total energy dissipation rate $Q_{\rm tot} (r)$
of the simulated turbulence (solid black curve). 
This rate is given by $Q_{\rm tot} = Q_{\perp} + Q_{\parallel}$, where
$Q_{\perp}$ is the contribution from damping at high perpendicular wavenumbers
(green curve), and $Q_{\parallel}$ is the contribution from damping at high parallel
wavenumbers (red curve), see equations (\ref{eq:Qperp}) and (\ref{eq:Qpara}).
Note that at large heights the parallel contribution is larger than the perpendicular
one; this is due to the weakness of the turbulence for the dominant waves in the
present model.
The dashed black curve shows the plasma heating rate $Q_{\rm A}$ used in the model
setup [same as the black curve in Figure \ref{fig1}(f)]. Figure \ref{fig2}(f) shows
the same heating rates per unit mass, $Q_{\rm tot}/\rho_0$ (full black curve) and
$Q_{\rm A}/\rho_0$ (dashed curve). Note that $Q_{\rm tot}/\rho_0$ has a minimum at
$r \approx 1.3$ $R_\odot$, near the height where the Alfv\'{e}n speed has its maximum
and the amplitude of the minority waves is reduced. Figures \ref{fig2}(e) and
\ref{fig2}(f) show that over a wide range of heights the dissipation rate
$Q_{\rm tot} (r)$ is significantly smaller than the plasma heating rate $Q_{\rm A} (r)$
needed to sustain the background atmosphere. We conclude that for the smooth model
considered here the simulated wave turbulence does not provide enough heating to
raise the temperature to the assumed level $T_0 (r)$ shown in Figure \ref{fig1}(c).

\subsection{Power Spectra and Wave Frequencies}

\begin{figure}
\epsscale{1.0}
\plotone{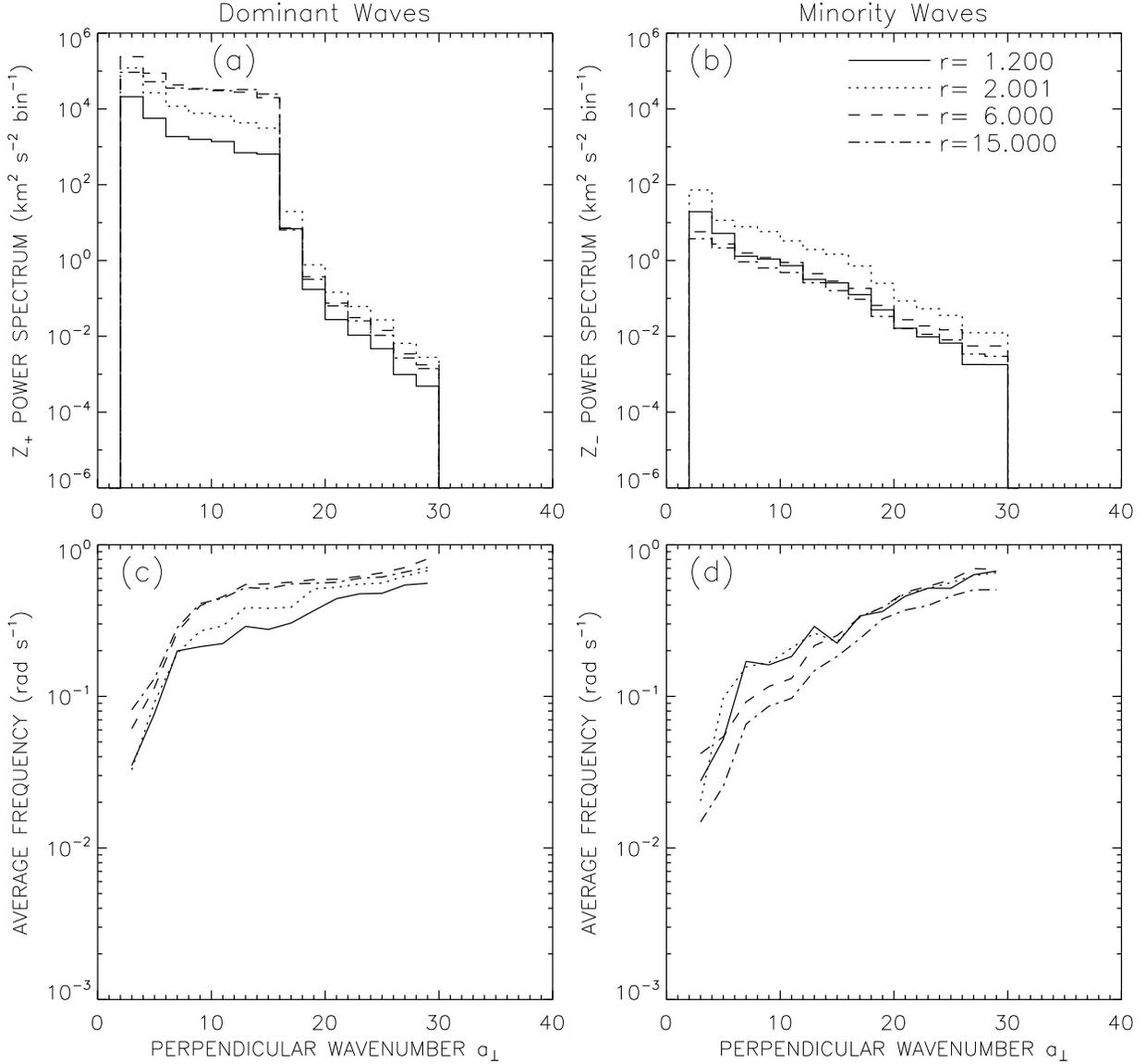}
\caption{Spatial power spectra and wave frequencies as function of
dimensionless perpendicular wavenumber $a_\perp$ for four different heights
in the smooth model. (a) Power spectra for $Z_{+}$, the Elsasser variable
for the dominant, outward propagating waves. The sharp drop at $a_\perp = 15$
is due to the onset of $\nu_k$-damping at that wavenumber. (b) Power spectra
for $Z_{-}$, the Elsasser variable for the minority waves, which also travel
outward. (c) Average wave frequencies for dominant waves. (d) Average wave
frequencies for minority waves. The different curves correspond to different
heights: $r = 1.2$ $R_\odot$ (solid), $r = 2$ $R_\odot$ (dotted),
$r = 6$ $R_\odot$ (dashed), $r = 15$ $R_\odot$ (dash-dotted).}
\label{fig3}
\end{figure}

Figures \ref{fig3}(a) and \ref{fig3}(b) show power spectra for the Elsasser
variables as function of dimensionless perpendicular wavenumber $a_\perp$ for
four different heights in the model. For each height we compute the wave power
in individual modes with wavenumbers $a_k$ ($k = 1, \cdots , 209$), and then
collect the results into bins in wavenumber space with $\Delta a_\perp = 2$
(for more details on how such spectra are computed, see section 4.2 of paper I).
These results are derived from the last 800 time steps of the simulation.
Figure \ref{fig3}(a) shows the power spectra for the outward waves. The first
and highest bin for each curve ($a_\perp \approx 3$) represents the outer scale
of the turbulence, and contains the driver modes ($a_k = 3.832$) that are
launched at the coronal base and propagate upward in height. The remaining
bins are filled by reflection-driven turbulent cascade. The sharp drop in
power at $a_\perp = 15$ is due to the onset of
damping $\nu_k$ at that wavenumber. At lower wavenumbers (where $\nu_k = 0$)
the power spectra are rather flat, dropping only about one order of magnitude
over the wavenumber range $2 < a_\perp < 15$. This flatness of the spectrum
may be due to a ``bottleneck effect" resulting from the use of a damping rate
$\nu_k$ that increases strongly with perpendicular wavenumber \citep[e.g.,][]
{Beresnyak2009b}. However, when considering the full wavenumber range the
power spectrum for outward waves is quite steep: the power drops by eight
orders of magnitude over the range $2 < a_\perp < 30$. Figure \ref{fig3}(b)
shows similar spectra for the minority waves, which have a much shallower
spectrum. Note that at low wavenumbers the minority wave power is much smaller
than the dominant wave power. For example, at $r = 6$ $R_\odot$ (dashed curve)
the power ratio is about $10^{-4}$, consistent with $Z_{-} / Z_{+} \sim 10^{-2}$
at that height in Figure \ref{fig2}(a).

We also compute {\it temporal} power spectra of dominant and minority waves,
and derive the average wave frequency $\tilde{\omega}_{\pm}$ as function of
dimensionless perpendicular wavenumber $a_\perp$. The results are shown
in Figures \ref{fig3}(c) and \ref{fig3}(d) for four different heights in
the model. At the outer scale of the turbulence ($a_\perp \approx 3$) the
outward waves are dominated by the driver modes. Using the model for random
footpoint motions described in Section 3, we find that the driver waves have
an average frequency $\tilde{\omega}_{+} \approx 2 \sqrt{\pi} \tau_0^{-1}
\approx 0.03$ $\rm rad ~ s^{-1}$, where $\tau_0$ (= 120 s) is the parameter
used in the setup of the random sequence. The left-most point on the curve
for $r = 1.2$ $R_\odot$ in Figure \ref{fig3}(c) is consistent with the
expected value of frequency for the driver waves. Note that the wave
frequencies $\tilde{\omega}_{\pm}$ generally increase with perpendicular
wavenumber $a_\perp$ for both dominant and minority waves.
Also, the frequencies of the dominant waves are somewhat larger than
the frequencies of the minority waves, even though both travel outward.

We now consider the question whether the turbulence in our model is weak
or strong. The turbulence is caused by nonlinear interactions between
Alfv\'{e}n waves \citep[][]{Iroshnikov1963, Kraichnan1965}. Following
\citet[][]{Chandran2009}, 
we consider the interactions between two wave packets ${\bf z}_{\pm}$
at the outer scale $\lambda_\perp$ of the turbulence, which we take to be
the inverse of the perpendicular wavenumber of the driver waves,
$\lambda_\perp = R / 3.832$.
The rate of shearing of the ${\bf z}_{\pm}$ wave packet by the ${\bf z}_{\mp}$
wave packet is given by $\tilde{\omega}_{\rm shear,\mp} = Z_{\mp} /
\lambda_\perp$. The waves interact for a certain ``collision" time, which
we take to be the inverse of the wave frequency in the comoving frame,
$t_{\rm coll, \mp} = 1 / \tilde{\omega}_{0,\mp}^\prime$, where the
subscript 0 indicates the outer scale. The nonlinearity parameter can be
defined as $\chi_{\mp} \equiv \tilde{\omega}_{\rm shear,\mp} t_{\rm coll, \mp}$.
If $\chi_{\mp} < 1$, then $\chi_{\mp}$ is approximately the fractional change
in the outer-scale ${\bf z}_{\pm}$ wave packet due to shearing by the
outer-scale ${\bf z}_{\mp}$ wave packet \citep[][]{Chandran2009}.
We find that for the ``smooth" model discussed in this Section,
the shearing rate $\tilde{\omega}_{\rm shear, -}$ relevant for the cascade of
the dominant waves varies from about $0.04$ $\rm rad ~ s^{-1}$ at $r = 1.1$
$R_\odot$ to less than $4 \times 10^{-4}$ $\rm rad ~ s^{-1}$ at $r > 10$
$R_\odot$.
At larger heights these shearing rates are significantly smaller than the
frequencies $\tilde{\omega}_{0,-}^\prime$ for the minority waves, which
determine the ``collision" time for dominant waves. In contrast, the shearing
rate $\tilde{\omega}_{\rm shear, +}$ relevant for the cascade of the minority
waves is comparable to the frequencies $\tilde{\omega}_{0,+}^\prime$
for the dominant waves. Therefore, in the present model the dominant outward
waves have large amplitudes but undergo weak turbulence ($\chi_{-} \ll 1$),
whereas the minority waves have small amplitudes but undergo strong turbulence
($\chi_{+} \sim 1$). This has important consequences for the wave dissipation
rate (see Section 5).

The power spectra shown in Figures \ref{fig3}(a) and \ref{fig3}(b) can be
compared with results from high-resolution simulations of {\it anisotropic}
MHD turbulence in a uniform background atmosphere. Here we focus on models
in which there is a large {\it imbalance} between counter-propagating
Alfv\'{e}n waves \citep[e.g.,][]{Beresnyak2008,Beresnyak2009a,
Beresnyak2009b, Perez2009, Perez2012}.
In such models the energy is injected by random forcing of the waves at
low perpendicular wavenumbers throughout the computational domain.
For example, \citet[][]{Perez2012} found that the spectra for the inertial
range are well fit by power laws with exponents of about -3/2 for the dominant
waves and slightly steeper for the subdominant waves. In contrast, in the
present model the background atmosphere is highly inhomogeneous, the outward
waves are launched at the coronal base, and minority waves are produced only by
wave reflections \citep[also see][]{Perez2013}. We find much steeper spectra
than in the homogeneous turbulence models, but this is likely due to the
relatively low spatial resolution of the model presented here. This is
confirmed by the work of \citet[][]{Perez2013}, who used higher resolution
and obtained spectra similar to those found in the homogeneous models.
Therefore, the spectra shown in Figures \ref{fig3}(a) and (b) are probably
not realistic. However, the main focus of our study is the wave dissipation
rate $Q_{\rm tot}$, and this quantity is likely to be much less affected
by limited spatial resolution.

\section{Model With Density Fluctuations}

The solar wind model discussed in Section 3 has the problem that the Alfv\'{e}n
wave dissipation rate $Q_{\rm tot} (r)$ predicted by the RMHD simulation is
much smaller than the heating rate $Q_{\rm A} (r)$ needed to sustain the
background atmosphere. Therefore, the model is not consistent from an energy
point of view. To obtain a more consistent model we must find a way to increase
$Q_{\rm tot}$ by about a factor $\sim 5$ without also increasing $Q_{\rm A}$.
In a reflection-driven wave-turbulence model the wave dissipation rate may be
increased by creating more wave reflection. We suggest that in the acceleration
region of the solar wind there are MHD waves of various type, not only Alfv\'{e}n
waves but also compressive, slow-mode waves traveling with velocities of the
order of the sound speed, $c_s \sim 100$ $\rm km ~ s^{-1}$. These sound waves
may in fact be produced by coupling with the Alfv\'{e}n waves \citep[e.g.,][]
{Kudoh1999, Moriyasu2004, Matsumoto2010}, and such coupling may also be involved
in the formation of jets and spicules \citep[][]{Tian2014, Cranmer2015b}. Sound waves
have associated density fluctuations, $\delta \rho (r,t)$, and since the magnetic
field $B_0 (r)$ is relatively unaffected, the Alfv\'{e}n speed $v_A (r,t)$ will
also vary in space and time. The Alfv\'{e}n waves reflect due to gradients in
Alfv\'{e}n speed, $d v_A /dr$ [see equation (\ref{eq:omeg_pm})], and in the
presence of density fluctuations these gradients may be significantly enhanced.
Therefore, sound waves with sufficient amplitude may act to ``scatter" the
dominant, outward propagating Alfv\'{e}n waves, producing more minority waves
and thereby enhancing the turbulent dissipation rate.

Observational evidence for density fluctuations comes from a variety of sources.
Radio observations have long been used to detect density fluctuations in the solar
wind \citep[e.g.,][]{Coles1989, Woo1996}. \citet[][]{Spangler2002} used radio
interferometry data to detect density variations with an amplitude of
$6 \%$ - $15 \%$ at heliocentric distances of 16-26 $R_\odot$. White light
eclipse images show a variety of coronal density structures \citep[e.g.,][]
{Druckmuller2014}. \citet[][]{Raymond2014} used the Atmospheric Imaging Assembly
(AIA) on the Solar Dynamics Observatory (SDO) to observe striations in the tail
of the sun-grazing Comet Lovejoy. These striations indicate the presence of large
density variations (at least a factor six) between neighboring coronal flux
tubes on scales of a few thousand kilometers. Most of these observations refer
to density variations {\it across} magnetic field lines. However, \citet[][]
{KrishnaPrasad2012} detected long-period intensity oscillations in open coronal
structures observed with AIA, and interpreted the results in terms of slow-mode
waves propagating {\it along} field lines. \citet[][]{Miyamoto2014} used spacecraft
radio occultation measurements at heights $1.5 < r < 20.5$ $R_\odot$, and found
quasi-periodic density disturbances with periods of 100 - 2000 s and amplitudes
of $30 \%$ at $r = 5$ $R_\odot$, which may also be due to slow-mode waves.
Using data from AIA, \citet[][]{Tian2011} found signatures of both longitudinal
and transverse waves in plume-like structures, rooted in magnetized regions of
the quiet solar atmosphere. The longitudinal waves have typical periods of
5 - 15 minutes and a phase speed of 120 $\rm km ~ s^{-1}$.
\citet[][]{Threlfall2013} compared wave observations from the Coronal Multi-channel
Polarimeter (CoMP) and AIA/SDO, and found evidence for transverse waves with
periods of 3 - 8 minutes and longitudinal waves with period of 6 - 11 minutes.
In a similar study, \citet[][]{Liu2015} found longitudinal waves with periods
of 10 - 20 minutes and a phase speed of 120 $\rm km ~ s^{-1}$. The associated
intensity oscillations have amplitudes of only $1 \%$, but this is likely due to
line-of-sight integration effects. A period of 15 minutes corresponds to a parallel
wavelength of about 0.16 $R_\odot$. Based on such measurements we assume density
variations along the magnetic field with an rms amplitude of $10 \%$ and an
auto-correlation length of 0.04 $R_\odot$, one quarter of the typical wavelengths
observed by \citet[][]{Tian2011} and \citet[][]{Liu2015}.

\begin{figure}
\epsscale{1.0}
\plotone{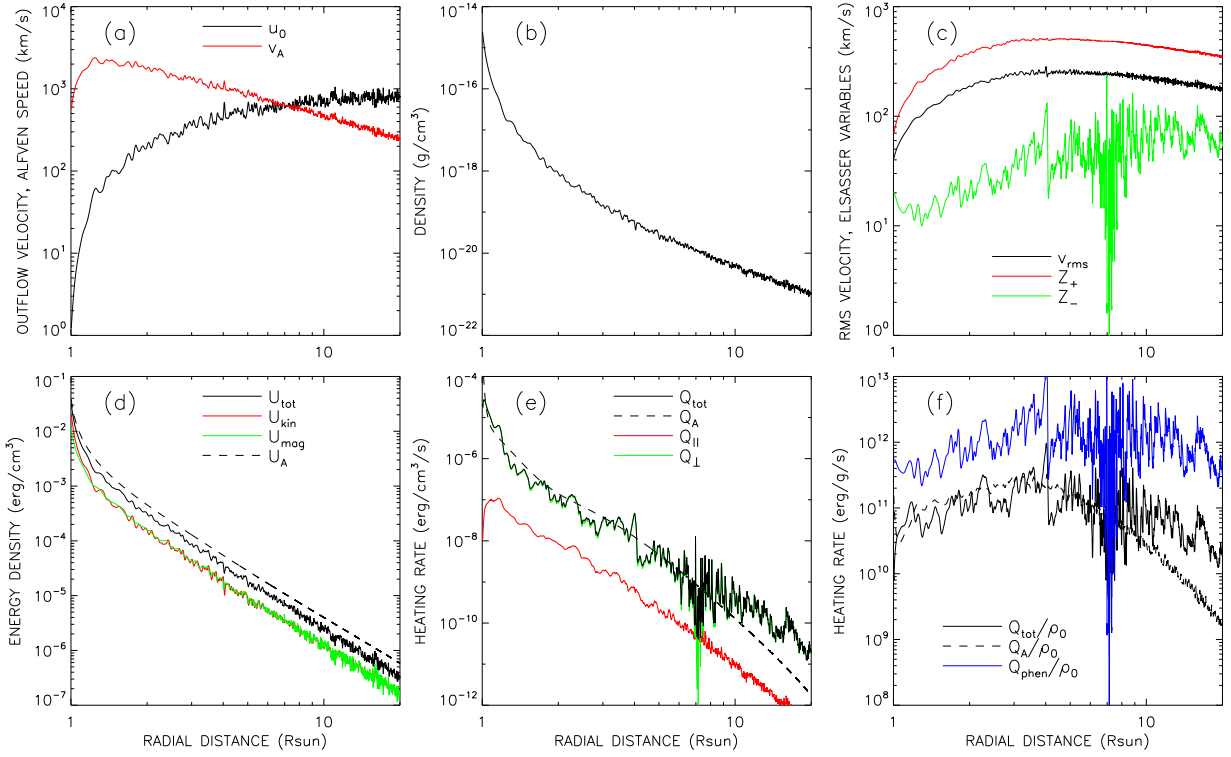}
\caption{Radial dependence of various quantities for a polar coronal hole
model with spatial variations in density along the flux tube. (a) Outflow
velocity (black curve) and Alfv\'{e}n speed (red curve). (b) Mass density.
(c) Velocity amplitude of the waves (black curve), and Elsasser variables
for the dominant waves (red curve) and minority waves (green curve).
In this model the minority waves have both inward- and outward-propagating
components.
(d) Wave energy densities as derived from the RMHD simulation: total energy
(black curve), kinetic energy (red curve), and magnetic energy (green curve).
Also shown is the wave energy density assumed in setup of the background
atmosphere (dashed curve). (e) Wave energy dissipation rates per unit volume:
total wave dissipation rate $Q_{\rm tot}$ (solid black curve), 
together with contributions from $Q_{\perp}$ (green curve) and
$Q_{\parallel}$ (red curve).
Also shown is the plasma heating rate $Q_{\rm A}$ assumed in setup of the
background atmosphere (dashed black curve).
(f) Wave energy dissipation rates per unit mass: rate derived from turbulence
simulation (solid black curve), rate assumed in the setup of background atmosphere
(dashed curve), and rate predicted by a phenomenological model, equation
(\ref{eq:Qphen3}) (blue curve).}
\label{fig4}
\end{figure}

In this section we consider a simple model for the effect of the density
fluctuations on the Alfv\'{e}n waves. For simplicity the density variations are
assumed to be {\it static}, i.e., independent of time. We construct a model
with spatial variations in density $\delta \rho (r)$ by taking the solution
from the ``smooth" model described in Section 3 and adding variations to certain
physical quantities. The resulting outflow velocity $u_0^\prime (r)$, Alfv\'{e}n
speed $v_A^\prime (r)$, and density $\rho_0^\prime (r)$ are shown in Figures
\ref{fig4}(a) and \ref{fig4}(b). The density is given by $\rho_0^\prime (r) =
\rho_0 (r) [ 1 + \epsilon (r) ]$, where $\rho_0 (r)$ is the density in the smooth
model, and $\epsilon (r) \equiv \delta \rho / \rho_0$ is a random function of
position (see below). The temperature $T_0 (r)$ and magnetic field strength
$B_0 (r)$ are assumed to be unaffected by the fluctuations, but the Alfv\'{e}n
speed $v_A^\prime (r) = v_A (r) [ 1 + \epsilon (r) ]^{-1/2}$, where $v_A (r)$
is the Alfv\'{e}n speed in the smooth model. To conserve mass and maintain the
same mass flux as in the smooth model, we assume that the outflow velocity
$u_0^\prime (r) = u_0 (r) / [ 1 + \epsilon (r) ]$. We make slight adjustments
to $\epsilon (r)$ near the Alfv\'{e}n critical point to ensure that this point
is crossed only once, i.e., there is only a single point $r_A$ in the model
where $u_0^\prime (r_A) = v_A^\prime (r_A)$. We found this adjustment is necessary
to avoid pile-up of ``inward" waves near the Alfv\'{e}n critical point.

The random function $\epsilon (r)$ is constructed as follows. We first create
a normally distributed random sequence $\epsilon^\prime (r)$ on a grid that is
uniform in radial distance. Then the sequence is Fourier filtered using a Gaussian
function $G(k_r) = \exp [ -(\lambda_0 k_r /2 \pi)^2 ]$, where $k_r$ is the radial
wavenumber, and $\lambda_0$ is a parameter that determines the correlation length
of the density variations. In this paper we take $\lambda_0 = 0.1$ $R_\odot$,
which corresponds to a correlation length $\lambda_{\rm c} = \lambda_0 /
\sqrt {2 \pi} \approx 0.04$ $R_\odot$. The filtered sequence is then renormalized
such that $\epsilon_{\rm rms} = 0.1$ and remapped onto the radial grid $r_n$
used for the numerical simulation. In this model both the correlation length
$\lambda_{\rm c}$ and fluctuation amplitude $\epsilon_{\rm rms}$ are assumed
to be independent of height.

The resulting one-dimensional model with spatial density variations is used
as the background atmosphere for three-dimensional RMHD simulations of the
Alfv\'{e}n waves. The boundary conditions and method of solution of the RMHD
equations are exactly the same as for the smooth model discussed in Section 3.
We find that in the model with density variations the minority waves have
both inward and outward-propagating components.
Figure \ref{fig4}(c) shows the simulation results for the rms velocity
amplitude of the waves (black curve) and the Elsasser variables for dominant
waves (red curve) and minority waves (green curve). These results
have been averaged over the cross-section of the flux tube and over time.
Note that the time-averaged Elsasser variable $Z_{-} (r)$ for the minority waves
shows strong spatial variations, especially near the Alfv\'{e}n critical point
where the ``inward" waves are nearly stationary. The mean value of $Z_{-} (r)$
is significantly increased compared to that in the smooth model [compare with
Figure \ref{fig2}(a)], and the dip in $Z_{-} (r)$ near $r = 1.3$ $R_\odot$ is
no longer present. This enhancement of the minority waves is due to additional
scattering that occurs in the model with density fluctuations. The Elsasser
variable $Z_{+} (r)$ for the dominant waves [red curve in \ref{fig4}(c)]
shows much smaller spatial variations, and the mean value is nearly unchanged
from that in the smooth model.

Figure \ref{fig4}(d) shows the total energy density $U_{\rm tot} (r)$ of
the simulated waves (solid black curve), together with the contributions from
kinetic energy (red curve) and magnetic energy (green curve). Note that
the fluctuations in these quantities are relatively small, which is due to
the fact that they are dominated by the outward-propagating waves.
The dashed curve in Figure \ref{fig4}(d) again shows the energy density
$U_{\rm A} (r)$ used in the setup of the background model [same as in Figure
\ref{fig1}(d)].

Figure \ref{fig4}(e) shows the energy dissipation rate $Q_{\rm tot} (r)$ as
derived from the RMHD simulations (solid black curve), together with
the contributions from $Q_{\perp}$ (green curve) and $Q_{\parallel}$
(red curve). These results are averaged over the cross-section of the flux
tube and over time. Note that the dissipation rates have large spatial variations
that are correlated with those of the Elsasser variable $Z_{-} (r)$ shown in
Figure \ref{fig4}(c). Comparing Figures \ref{fig2}(e) and \ref{fig4}(e),
we see that the dissipation rate $Q_{\rm tot} (r)$ is significantly increased
compared to its value in the smooth model. The dashed curve in Figure
\ref{fig4}(e) shows the plasma heating rate $Q_{\rm A} (r)$ needed to sustain
the background atmosphere. We see that the mean value of $Q_{\rm tot} (r)$
approximately equals the heating rate $Q_{\rm A} (r)$ at all heights up to
$r = 8$ $R_\odot$, and exceeds $Q_{\rm A}$ at larger heights, i.e., {\it in the
model with density fluctuations the simulated wave turbulence produces enough
energy dissipation to heat the background atmosphere}. The same can be seen in
Figure \ref{fig4}(f), where we plot the dissipation rate per unit mass,
$Q_{\rm tot} / \rho_0^\prime$ (solid black curve), and the heating rate per
unit mass, $Q_{\rm A} / \rho_0^\prime$ (dashed black curve). For heights in
the range 1 to 10 $R_\odot$ the mean dissipation rate is about $10^{11}$
$\rm erg ~ g^{-1} ~ s^{-1}$, similar to the values found in the models by
\citet[][]{Cranmer2007} (see their Figure 7) and \citet[][]{Chandran2011}
(their Figure 3b). However, these earlier models did not include the effects
of density fluctuations on the reflection of the Alfv\'{e}n waves.

\section{Phenomenology for Turbulence in an Inhomogeneous Atmosphere}

Several authors have developed ``phenomenological" models for the dissipation
rate $Q$ in homogeneous MHD turbulence. For decaying turbulence the dissipation
rate may be approximated as
\begin{equation}
Q_{\rm phen} = \rho_0 \frac{Z_{+}^2 Z_{-} + Z_{-}^2 Z_{+}}
{4 \lambda_\perp} ,  \label{eq:Qphen1}
\end{equation}
where $Z_{\pm}$ are the rms values of the Elsasser variables, and
$\lambda_\perp$ is the outer scale of the turbulence \citep[][]{Hossain1995}.
The same expression has been applied for reflection-driven turbulence in the
solar wind \citep[e.g.,][]{Zhou1990, Matthaeus1999, Dmitruk2001, Dmitruk2002}.
\citet[][]{Dmitruk2003} found good agreement between the above phenomenological
model and results from numerical RMHD simulations. \citet[][]{Cranmer2007} and
\citet[][]{Chandran2011} constructed detailed models of the fast solar wind based
on such expressions for the wave dissipation rate.

\begin{figure}
\epsscale{1.0}
\plotone{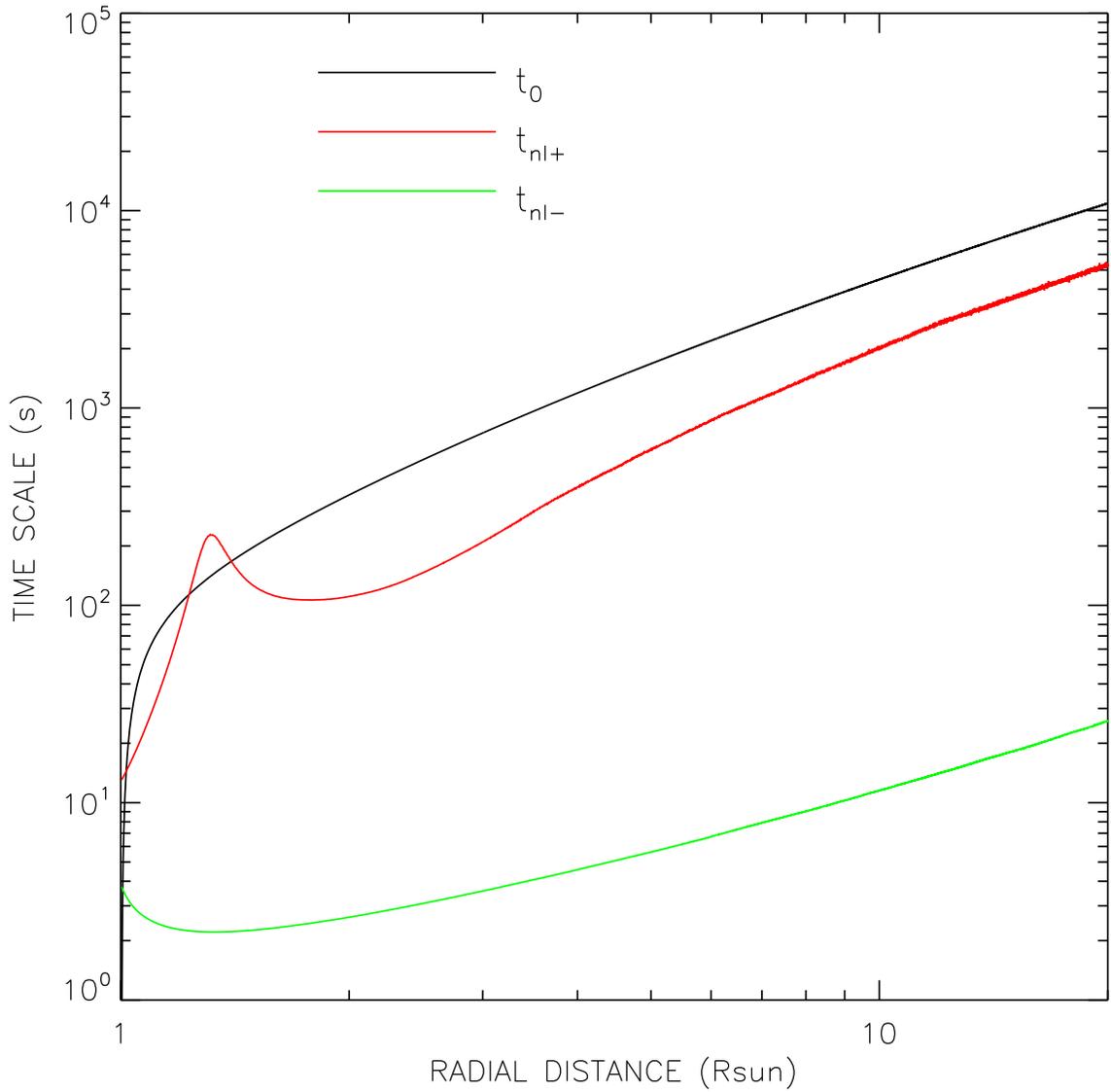}
\caption{Various time scales as function of radial distance $r$ in the model
with a smooth background atmosphere. The red and green curves show the nonlinear
times $t_{\rm nl,\pm} (r)$ for the dominant and minority waves, respectively.
The black curve shows the time $t_0 (r)$ for an outward propagating wave to reach
a certain height $r$.}
\label{fig5}
\end{figure}

Equation (\ref{eq:Qphen1}) is based on the assumption that the cascade times for
dominant and minority waves are given by the nonlinear time scales, $t_{\rm nl,\pm}
\equiv \lambda_\perp / Z_{\mp}$.
\citet[][]{Chandran2009} consider reflection-driven turbulence and argue that
these expressions are appropriate when the dominant waves undergo weak turbulence
($\chi_{-} \ll 1$) and the minority waves undergo strong turbulence ($\chi_{+} \sim 1$),
as is the case in our ``smooth" model (Section 3). \citet[][]{Perez2009} and
\citet[][]{Mallet2015} use a different expression for the nonlinear time that
also includes a dependence on the alignment angle $\theta$ between the
${\bf z}_{+}$ and ${\bf z}_{-}$ vectors; when this angle is small, the nonlinear
coupling between the waves is further reduced, lengthening the nonlinear times.
\citet[][]{Perez2013} found evidence for such alignments in models with long
correlations times, but we find no evidence for such alignments in our simulations.
Therefore, we omit the dependence on $\theta$ in the above definition of the
nonlinear times. Figure \ref{fig5} shows the nonlinear times $t_{\rm nl,\pm}$
(red and green curves) for the ``smooth" model discussed in Section 3. The black
curve shows the wave travel time $t_0 (r)$, i.e., the time for an outward-propagating
wave to travel from the coronal base to a specific radial distance $r$. Note that
the nonlinear time for the dominant outward waves is comparable to the wave travel
time, $t_{\rm nl,+} \sim t_0$. Therefore, the dominant waves do not have time to
efficiently develop a turbulent spectrum before they escape into the region
$r > 20$ $R_\odot$. This explains why the power spectrum for these waves deviates
significantly from a power law, see Figure \ref{fig3}(a). In contrast,
the nonlinear time for the minority waves is much smaller than $t_0 (r)$, so the
turbulence is well developed for the minority waves.

In the present work we find that equation (\ref{eq:Qphen1}) significantly
overestimates the dissipation rate compared to the value $Q_{\rm tot} (r)$ derived
from our numerical simulations. For example, at $r = 2$ $R_\odot$ in the ``smooth"
model equation (\ref{eq:Qphen1}) overestimates the actual dissipation rate by about
a factor 100. One possible reason for the discrepancy might be that the present numerical
modeling is somehow deficient and severely underestimates the actual dissipation rate.
It is true that the spatial resolution of our RMHD simulations
is not very high, and our assumption of a flux tube with rigid boundary is questionable.
However, we do not believe these effects can cause the dissipation rate $Q_{\rm tot}$
to be underestimated by such a large factor. Another possibility is that equation
(\ref{eq:Qphen1}) may not be applicable to our simulation, perhaps because one or
more of the assumptions behind the equation are not valid for our case. However,
we have not been able to identify any reason why the equation would not be applicable;
the arguments by \citet[][]{Chandran2009} in favor of this expression would seem to be
valid in our case. Therefore, it is unclear why equation (\ref{eq:Qphen1}) gives such
a poor fit to the numerically computed dissipation rate.

We tried other formulae for $Q_{\rm phen}$ in an attempt to obtain a better fit.
Since the first term in equation (\ref{eq:Qphen1}) is much larger than the second term,
let us assume that the first term is somehow reduced by a factor ${\cal E}_{+} < 1$.
Then equation (\ref{eq:Qphen1}) can be generalized as follows:
\begin{equation}
Q_{\rm phen} = \rho_0 \frac{{\cal E}_{+} Z_{+}^2 Z_{-} 
+ Z_{-}^2 Z_{+}} {4 \lambda_\perp} .  \label{eq:Qphen2}
\end{equation}
The largest reduction is obtained when ${\cal E}_{+} \sim Z_{-}/ Z_{+}$, so that
the dominant and minority waves have approximately equal contributions to the energy
dissipation rate (further reduction of ${\cal E}_{+}$ would have only a minor effect).
Assuming exact equality of the two contributions (${\cal E}_{+} = Z_{-}/ Z_{+}$),
we obtain
\begin{equation}
Q_{\rm phen} = \rho_0 \frac{Z_{-}^2 Z_{+}} {2 \lambda_\perp} .
\label{eq:Qphen3}
\end{equation}
We used this expression to compute $Q_{\rm phen} (r)$ for both the ``smooth" model
and the model with density fluctuations. The blue curve in Figure \ref{fig2}(f) shows
$Q_{\rm phen}/ \rho_0$ for the ``smooth" model discussed in Section 3. Note that at
low heights equation (\ref{eq:Qphen3}) still overestimates the numerically computed
rate $Q_{\rm tot}$ (solid back curve); for example, at $r = 2$ $R_\odot$ the ratio
$Q_{\rm phen} / Q_{\rm tot} \approx 4$. Although this is not a good fit, the ratio
is much smaller than that obtained with equation (\ref{eq:Qphen1}), which predicts
$Q_{\rm phen} / Q_{\rm tot} \approx 100$. For $r > 6$ $R_\odot$ equation
(\ref{eq:Qphen3}) underestimates the numerically computed rate, but when $Q_{\rm phen}$
is compared with only the perpendicular contribution $Q_\perp$ the agreement with the
numerical results is significantly improved. 
The blue curve in Figure \ref{fig4}(f) shows the phenomenological rate
$Q_{\rm phen} / \rho_0^\prime$ for the model with density fluctuations (Section 4).
In this case equation (\ref{eq:Qphen3}) overestimates $Q_{\rm tot} (r)$ by a factor
ranging from 7 to 20. We conclude that equation (\ref{eq:Qphen3}) provides a much
better fit to the data than equation (\ref{eq:Qphen1}), but still shows significant
discrepancies between $Q_{\rm phen}$ and $Q_{\rm tot}$.

It should be mentioned that equation (\ref{eq:Qphen3}) depends on the rms
amplitudes of the dominant and minority waves, $Z_{\pm} (r)$. In the present work
we were able to derive these amplitudes from the numerical RMHD simulations.
However, for modeling the solar wind in the manner of \citet[][]{Cranmer2007}
and \citet[][]{Chandran2011} it would be useful to obtain accurate
approximations for $Z_{\pm} (r)$, so that the modeling can be done without
doing computationally intensive RMHD simulations. Developing such approximations
for models with density fluctuations is not trivial, and is beyond the scope
of the present project. However, we realize that without such approximations
the above equations for $Q_{\rm phen}$ are of limited use.

\section{Discussion and Conclusions}

In this paper we considered a simple, one-fluid model of the fast solar wind,
and we neglected all details of the collisionless processes by which the waves
are dissipated at small spatial scales. In reality the solar wind exhibits
significant departures from thermal equilibrium: different particle species have
different temperatures, and particle velocity distributions can deviate significantly
from Maxwellian \citep[e.g.,][]{Feldman1997, Kohl2006}. When Alfv\'{e}n wave energy
cascades to the proton gyro-radius scale $\rho_p $, some of the energy may be
dissipated by linear and nonlinear damping at that scale ($k_\perp \rho_p
\approx 1$), and the remainder of the energy may cascade into the kinetic
Alfv\'{e}n wave regime ($k_\perp \rho_p \gg 1$), where the damping mainly benefits
the electrons \citep[see][and references therein]{Chandran2011}. Therefore, the
partitioning of the energy between ions and electrons depends on the details of
these linear and nonlinear processes at and below the proton gyro-radius scale.
In this paper we assume that the total dissipation rate is insensitive to the
details of the dissipation process.

In Section 3 we found that in the model with a smooth background atmosphere
the reflection-driven turbulence does not provide enough heating to maintain
the assumed temperature, $T_0 (r)$. We explored other values of the model
parameters, and found that if the temperature is reduced [by using $C_0 = 0.3$
in equation (\ref{eq:T0})], the wave pressure force becomes even more dominant,
and the outflow velocity at $r = 20$ $R_\odot$ increases beyond 1000
$\rm km ~ s^{-1}$, too high for a realistic model of the fast solar wind.
On the other hand, if the temperature is raised ($C_0 = 0.4$), the required
heating rate $Q_{\rm A} (r)$ increases in the central part of the model
($2 R_\odot < r < 10 R_\odot$), and the wave action parameter $S_{\rm A} (r)$
is reduced by 90\% over the height range of the model, which is also not
realistic \citep[see, however,][]{Hahn2012}. If the wave amplitude at the
coronal base is reduced from 40.6 $\rm km ~ s^{-1}$ to 29.5 $\rm km ~ s^{-1}$,
the wave action parameter even becomes negative, so there is not enough energy
to heat the plasma at larger heights. In all three cases the dissipation rate
$Q_{\rm tot} (r)$ remains well below the heating rate $Q_{\rm A} (r)$ needed
in the central part of the model. Hence, there does not appear to be a smooth
background atmosphere for which the turbulence can provide enough heating.

In Section 4 we considered the effects of density fluctuations on the
propagation and reflection of the Alfv\'{e}n waves. Such fluctuations may be
due to compressive waves in the solar wind \citep[][]{Kudoh1999, Moriyasu2004, 
Matsumoto2010}. We found that density variations with an rms amplitude of
$10 \%$ and correlation length 0.04 $R_\odot$ produce strong wave reflections
that significantly enhance the amplitude of the minority waves,
and thereby the wave dissipation rate. The time-averaged wave dissipation
rate is approximately equal to the plasma heating rate needed to maintain
the temperature of the background atmosphere, i.e., the model with density
fluctuations is approximately in thermal equilibrium. This suggest that
Alfv\'{e}n wave turbulence can heat and accelerate the fast solar wind,
provided the effects of density fluctuations on wave reflection are taken
into account.

In Section 5 we compared our simulation results with predictions from
``phenomenological" turbulence models, taking into account the strong imbalance
between dominant and minority waves ($Z_{+} \gg Z_{-}$). We found that the
standard formula for the energy dissipation rate, equation (\ref{eq:Qphen1}),
significantly overestimates the numerically computed rate for the model with
a smooth background atmosphere. The reasons why this formula gives such a poor
fit are not fully understood. We proposed a revised formula based on the
assumption that the cascade rate for the dominant waves is significantly
reduced. We found that this revised formula [equation (\ref{eq:Qphen3})]
provides a better fit to the numerically computed rate, although there are
still significant discrepancies. \citet[][]{Cranmer2007} and
\citet[][]{Chandran2011} used the standard formula to construct detailed models
of the solar wind, neglecting the effects of density fluctuations. We suggest
that these authors may have overestimated the wave heating rate.

The ratio $Z_{-}/Z_{+}$ is about a factor 10
larger in the model with density fluctuations than in the smooth model, so
this quantity could be an important indicator for the presence of density
fluctuations. \citet[][]{Bavassano2000} used {\it Ulysses} observations to
determine the energy densities of outward- and inward-propagating waves as
function of radial distance in the heliosphere. Combining their results with
{\it Helios} observations and extrapolating back to $r = 0.1$ AU, they find
$Z_{-}/Z_{+} \sim 0.1$, similar to the value in our model with density
fluctuations [see Figure \ref{fig4}(c)]. However, the waves observed in the
heliosphere have periods of about 1 hour, much longer than the periods of
the waves simulated here. Therefore, these heliosphere observations do not
provide strong constraints on the present modeling. Such long-period waves
may be produced by an inverse cascade of wave energy to large perpendicular
scales. Since we consider only a single, relatively narrow flux tube
($R_{\rm base} = 1$ Mm), and do not include interactions between neighboring
flux tubes, such a cascade cannot be described with the present model.

\citet[][]{Morton2015} presented observational evidence for 
inward-propagating Alfv\'{e}n waves in coronal holes, based on Dopplergrams
obtained with the CoMP instrument. According to Figure 3 of their paper,
the ratio of power spectra for inward and outward waves with frequencies in
the range 3 - 7 mHz is about 0.4, much larger than the value of about 0.02
predicted by the present models [ratio $(Z_{-}/Z_{+})^2$ derived from
the red and green curves in Figure \ref{fig2}(a) or Figure \ref{fig4}(c) for
$r \approx 1.05$ $R_\odot$]. In our model the level of minority waves near
the coronal base is mainly determined by the rapid rise in Alfv\'{e}n speed
from about 700 $\rm km ~ s^{-1}$ at the coronal base ($r = 1.003$ $R_\odot$)
to about 2000 $\rm km ~ s^{-1}$ at $r \approx 1.3$ $R_\odot$, which causes
wave reflection. In contrast, the observations show a nearly constant wave
propagation speed of about 400 $\rm km ~ s^{-1}$, which should produce less
reflection than predicted by our model. This suggests that the observed
inward waves are mainly produced by small-scale density fluctuations, not by
reflections due to the overall height dependence of the mean Alfv\'{e}n
speed $v_A (r)$. Further observations of longitudinal and transverse waves
in coronal holes would be very useful in clarifying the origin of the inward
waves, and for constraining the type of models developed in this paper.

Figure \ref{fig4}(c) shows large fluctuations in the rms value of the
Elsasser variable $Z_{-} (r)$ for the minority waves, even
though the simulation results have been averaged over the cross-section
of the flux tube and over time. The large fluctuations are likely an
artifact of our assumption that the density variations $\delta \rho (r)$
are static, independent of time. In reality the density is expected to
fluctuate in space and time, and we speculate that including the effects
of temporal variability will reduce the magnitude of the fluctuations in
$Z_{-} (r)$, but will not affect the mean value of $Z_{-}$, which will
still be enhanced compared to a model without density fluctuations.
However, this hypothesis cannot be tested with the present RMHD model,
which assumes a fixed background atmosphere.

The present modeling still neglects any variations in density over the
cross-section of the flux tube. However, large density variations across
field lines have been observed \citep[e.g.,][]{Woo1996, Spangler2002,
Raymond2014, Tian2011, Threlfall2013, Liu2015}, and such variations likely
have a significant effect on the propagation of Alfv\'{e}n waves.
To simulate the effects of perpendicular density variations on the Alfv\'{e}n
waves will require full MHD modeling. Such modeling is also needed to
account for the coupling between Alfv\'{e}n waves and other types of MHD waves.
Previous studies have shown that density variations in the perpendicular
direction can drastically change the nature of the waves, and can lead
to phase-mixing and resonant absorption of the waves \citep[e.g.,][]
{Heyvaerts1983, DeGroof2002, Goossens2011, Goossens2012, Goossens2013,
Pascoe2012}. Future modeling of the fast solar wind using full MHD
simulations should take such effects into account.

\acknowledgements 
We thank the referee for providing detailed comments that helped improve the
paper. We are most grateful to Alex Voss from the School of Computer Science
at the University of St.~Andrews for his support with the computational work.
We thank Steve Cranmer for his thorough reading of the manuscript and helpful
comments. We also thanks Hui Tian for pointing out recent observations relevant
to our work. We are grateful to Benjamin Chandran for providing more information
about the modeling results of \citet[][]{Perez2013}.
This project was supported under contract NNM07AB07C from NASA to the Smithsonian
Astrophysical Observatory (SAO) and SP02H1701R from LMSAL to SAO. This research
has made use of NASA's Astrophysical Data System.

\clearpage

\appendix

\section{Reduced MHD Model for the Solar Wind}

The RMHD equations are a simplified version of the full MHD equations
(\ref{eq:mass}), (\ref{eq:dvdt}), (\ref{eq:heat}) and (\ref{eq:dBdt}). Actually,
RMHD involves several approximations: (1) the magnetic fluctuations associated
with the waves are assumed to have a transverse length scale $\ell_\perp$ that
is small compared to their parallel scale $\ell_\parallel$; (2) the amplitude
of the magnetic fluctuations is assumed to be small compared to the background
field, $| {\bf B}_1 | \ll B_0$; (3) the velocity fluctuations are assumed to be
small compared to the Alfv\'{e}n speed, $| {\bf v}_1 | \ll v_A$; (4) the plasma
pressure and density are assumed to be equal to their background values,
$p \approx p_0$ and $\rho \approx \rho_0$, i.e., we neglect the coupling of the
Alfv\'{e}n waves with compressive, slow- and fast-mode waves; (5) the background
density and field strength are assumed to be constant over the cross-section
of the flux tube. Then the MHD equations can be split into two sets
of coupled equations, one for the background medium and another for the Alfv\'{e}n
waves. In paper I we presented a detailed derivation of the RMHD equations for
the case where the effects of parallel flows on the waves can be neglected,
$u_0 \ll v_A$. However, this approximation is not valid for the solar wind
because $u_0 = v_A$ at the Alfv\'{e}n critical point, which is located at
$r \sim 10$ $R_\odot$ \citep[][]{Cranmer2007, Perez2013}. Therefore,
we now consider the effect of $u_0$ on the dynamics of the waves.

As in paper I, the background magnetic field ${\bf B}_0 ({\bf r})$ is assumed
to be a potential field, $\nabla \times {\bf B}_0 = 0$, and is locally
approximated as
\begin{equation}
{\bf B}_0 (x,y,r) \approx B_0 \hat{\bf r} - \frac{1}{2} \frac{dB_0} {dr}
( x \hat{\bf x} + y \hat{\bf y} ) , \label{eq:B0}
\end{equation}
where $r$ is the coordinate along the flux tube axis, $x$ and $y$ are coordinates
perpendicular to the axis, $B_0 (r)$ is the field strength on axis, and
$\hat{\bf r}$, $\hat{\bf x}$ and $\hat{\bf y}$ are unit vectors.
In this paper the flux tube is assumed to be radially oriented, but this is not
essential for the equations described in this Appendix. The unit vector
$\hat{\bf B}_0$ along the background field varies over the cross-section of the
tube, and is given by
\begin{equation}
\hat{\bf B}_0 (x,y,r) \approx \hat{\bf r} - \frac{1}{2 H_{\rm B}}
( x \hat{\bf x} + y \hat{\bf y} ) ,  \label{eq:B0hat}
\end{equation}
where $H_{\rm B} (r) \equiv B_0 / (dB_0 /dr)$ is the length scale for variations
of the background field ($H_{\rm B} < 0$). The radius $R(r)$ of the tube is 
assumed to be small compared to $| H_{\rm B} |$. The Alfv\'{e}n waves cause
perturbations of the magnetic field ${\bf B} ({\bf r},t)$ inside the tube.
The induction equation (\ref{eq:dBdt}) can be written in the form:
\begin{equation}
\frac{\partial {\bf A}} {\partial t} = {\bf v} \times {\bf B} + \nabla \phi
+ {\bf D}_A ,  \label{eq:dAdt}
\end{equation}
where ${\bf A} ({\bf r},t)$ is the vector potential (${\bf B} \equiv \nabla \times
{\bf A}$), $\phi ({\bf r},t)$ is a scalar potential, ${\bf v} ({\bf r},t)$ is the
plasma velocity, and ${\bf D}_A$ is a dissipative term. The velocity field is
approximated as
\begin{equation}
{\bf v} ({\bf r},t) \approx u_0 \hat{\bf B}_0 +
\nabla_\perp f \times \hat{\bf B}_0 , \label{eq:v1}
\end{equation}
where $u_0 (r)$ is the outflow velocity of the solar wind, $f({\bf r},t)$ is the
stream function of the velocity perturbations, and $\nabla_\perp$ is the derivative
perpendicular to the background field, $\nabla_\perp \equiv \nabla - \hat{\bf B}_0
( \hat{\bf B}_0 \cdot \nabla )$. Following \citet[][]{Strauss1997}, we assume that the
first-order perturbation of the vector potential ${\bf A}_1$ is parallel to the
background field:
\begin{equation}
{\bf A}_1 ({\bf r},t) \approx h({\bf r},t) {\bf B}_0 ({\bf r}) ,  \label{eq:A1}
\end{equation}
where $h({\bf r},t)$ is the magnetic flux function. Since $\nabla \times {\bf B}_0
= 0$, it follows that the perturbed magnetic field can be approximated as
\begin{equation}
{\bf B} ({\bf r},t) \approx {\bf B}_0 + \nabla_\perp h \times {\bf B}_0 .
\label{eq:B1}
\end{equation}
Therefore, the cross-product of ${\bf v}$ and ${\bf B}$ is given by
\begin{equation}
{\bf v} \times {\bf B} \approx u_0 B_0 \nabla_\perp h - B_0 \nabla_\perp f 
+ B_0 \left[ \hat{\bf B}_0 \cdot ( \nabla_\perp f \times \nabla_\perp h ) \right]
\hat{\bf B}_0 ,  \label{eq:vxB}
\end{equation}
and inserting this into equation (\ref{eq:dAdt}), we obtain for the parallel and
perpendicular components of this equation:
\begin{eqnarray}
\frac{\partial h}{\partial t} & \approx & \frac{1}{B_0} \hat{\bf B}_0 \cdot \nabla \phi
+ \hat{\bf B}_0 \cdot ( \nabla_\perp f \times \nabla_\perp h ) , 
\label{eq:dhdt0} \\
0 & \approx & u_0 B_0 \nabla_\perp h - B_0 \nabla_\perp f + \nabla_\perp \phi .
\label{eq:nabla_phi}
\end{eqnarray}
The latter can be integrated over $x$ and $y$ to yield an expression for $\phi$,
and inserting this expression into equation (\ref{eq:dhdt0}) yields
\begin{equation}
\frac{\partial h} {\partial t} = \hat{\bf B}_0 \cdot \nabla ( f - u_0 h ) 
+ \frac{f - u_0 h} {H_{\rm B}} + [f,h] + D_h , \label{eq:dhdt}
\end{equation}
where $D_h$ is a dissipative term. Here the bracket operator is defined by
\begin{equation}
[ a,b ] \equiv \frac{\partial a} {\partial x} \frac{\partial b} {\partial y} -
\frac{\partial a} {\partial y} \frac{\partial b} {\partial x} ,
\label{eq:bracket}
\end{equation}
where $x$ and $y$ are the coordinates perpendicular to the flux tube axis.
All nonlinearities of the RMHD model are contained within such bracket terms.

A similar analysis can be applied to the equation of motion (\ref{eq:dvdt}).
The perpendicular component of this equation yields
\begin{equation}
\left( \frac{d {\bf v}} {dt} \right)_\perp = \frac{u_0}{R} \hat{\bf B}_0
\cdot \nabla (R {\bf v}_1 ) + \frac{\partial {\bf v}_1} {\partial t} +
{\bf v}_1 \cdot \nabla {\bf v}_1 , \label{eq:dvdt_perp}
\end{equation}
where we used equation (\ref{eq:B0hat}), and we assumed flux conservation
($B_0 R^2$ = constant). By taking the curl of equation (\ref{eq:dvdt_perp}),
we obtain the following vorticity equation:
\begin{equation}
\frac{\partial \omega} {\partial t} = 
- u_0 \left( \hat{\bf B}_0 \cdot \nabla \omega  - \frac{\omega} {H_{\rm B}} \right)
- [\omega, f] + v_A^2 \left\{ \hat{\bf B}_0 \cdot \nabla \alpha + [\alpha, h]
\right\} + D_\omega , \label{eq:dodt}
\end{equation}
where $\alpha ({\bf r},t) \equiv - \nabla_\perp^2 h$ is the magnetic torsion
parameter, $\omega ({\bf r},t)$ is the parallel component of vorticity:
\begin{equation}
\omega ({\bf r},t) \equiv \hat{\bf B}_0 \cdot \nabla \times {\bf v}_1
\approx - \nabla_\perp^2 f ,
\end{equation}
and $D_\omega$ is a dissipative term. In deriving equation (\ref{eq:dodt})
we neglected terms of higher order in $\epsilon$ ($\equiv \ell_\perp /
\ell_\parallel$), and we used equation (\ref{eq:B0hat}) to compute the
$x$- and $y$-derivatives of $\hat{\bf B}_0$. The first term in
equation (\ref{eq:dodt}) describes the torque due to the expansion of the
plasma in the parallel flow $u_0$. For a detailed derivation of the other
terms, see paper I. The RMHD approximations greatly simplify the MHD equations,
reducing them to two coupled equations (\ref{eq:dhdt}) and (\ref{eq:dodt})
for two scalar quantities, $h({\bf r},t)$ and $f({\bf r},t)$. The key feature
of the RMHD equations is that they retain the nonlinear terms responsible for
the development of Alfv\'{e}n wave turbulence. Note that the outflow velocity
$u_0 (r)$ only affects the linear terms in the equations.

\section{Numerical Methods}

In this paper we consider Alfv\'{e}n waves propagating along a thin flux tube
with {\it circular} cross-section. The radius $R(r)$ of the cross-section
increases with distance $r$ along the tube. For an arbitrary point within the
tube, let $\tilde{r} \equiv \sqrt{x^2 + y^2} \le R(r)$ be the distance from
the axis, and let $\varphi$ be the azimuth angle. Then the scalar functions
can be written as $h(\xi,\varphi,r,t)$ and $f(\xi,\varphi,r,t)$,
where $\xi \equiv \tilde{r}/R$ is the fractional distance from the flux tube
axis, and derivatives along the background field can be written as partial
derivatives $\partial / \partial r$ at constant $\xi$ and $\varphi$.
We use a spectral method to describe the dependence of $h$ and $f$ on the
perpendicular coordinates $\xi$ and $\varphi$. Specifically, we use a set
of orthogonal basis functions $F_k (\xi,\varphi)$ that are eigenmodes of
the $\nabla_\perp^2$ operator and also satisfy the side boundary conditions
on the flux tube (see Appendix B of paper I). The modes are enumerated by
an index $k$ ($k = 1, \cdots , N$) and have well-defined perpendicular
wavenumbers $k_\perp = a_k/R$, where $a_k$ is a dimensionless wavenumber
(given by the zeros of Bessel functions). For the simulations presented in
this paper, the maximum dimensionless wavenumber $a_{\rm max} = 30$, which
requires $N = 209$ modes.

The magnetic and velocity fluctuations are described by the mode amplitudes
$h_k (r,t)$ and $f_k (r,t)$, respectively. The RMHD equations can then be
written as
\begin{eqnarray}
\frac{\partial h_k} {\partial t} & = & \frac{\partial} {\partial r}
( f_k - u_0 h_k ) + \frac{ f_k - u_0 h_k } {H_{\rm B}} 
+ \frac{1}{R^2} \sum_{j=1}^N \sum_{i=1}^N M_{kji} f_j h_i  \nonumber \\
 & & - \nu_k h_k + \beta D^6 h_k , \label{eq:dhdt1} \\
\frac{\partial \omega_k} {\partial t} & = & - u_0 \left(
\frac{\partial \omega_k } {\partial r} - \frac{\omega_k} {H_{\rm B}} \right)
+ v_A^2 \frac{\partial \alpha_k } {\partial r} 
+ \frac{1}{R^2} \sum_{j=1}^N \sum_{i=1}^N M_{kji} \left( v_A^2
\alpha_j h_i - \omega_j f_i \right)  \nonumber \\
 & & - \nu_k \omega_k + \beta D^6 \omega_k , \label{eq:dodt1}
\end{eqnarray}
where $\alpha_k = (a_k /R)^2 h_k$ and $\omega_k = (a_k /R)^2 f_k$ are the
mode amplitudes for magnetic torsion and vorticity, and $M_{kji}$ is an
anti-symmetric matrix describing the nonlinear couplings between the various
modes (see paper I). Here we added artificial damping terms involving the
parameters $\nu_k$ and $\beta$. The damping rate $\nu_k$ depends on
the dimensionless {\it perpendicular} wavenumber $a_k$ of the waves. 
Note that the same damping $\nu_k$ is applied to both the magnetic-
and velocity fluctuations, so the magnetic Prandtl number $Pr_{\rm M} = 1$.
For low perpendicular wavenumbers ($a_k \le 15$) we set $\nu_k = 0$, so that
the outward propagating waves can travel to large height without any damping.
For high wavenumbers ($15 \le a_k \le 30$) the damping rate increases linearly
with $a_k$, and reaches its maximum value $\nu_{\rm max}$ at $a_k = 30$.
The maximum rate is given by $\nu_{\rm max} (r,t) = 70 ~ \overline{v_{\rm rms}}
(r,t) / R(r)$, where $v_{\rm rms} (r,t)$ is the rms velocity of the waves,
and the bar denotes a running time average over a time interval of 2000 s.
The terms with $\beta$ in equations (\ref{eq:dhdt1}) and (\ref{eq:dodt1})
involve the sixth power of the dimensionless derivative operator $D \equiv
(u_0 + v_A ) \Delta t \partial / \partial r$. The purpose of these terms is
to prevent the build-up of waves with high {\it parallel} wavenumbers that
cannot be adequately resolved on the radial grid; we use $\beta = 0.001 /
(64 \Delta t)$. The terms with $\nu_k$ and $\beta$ represent the physical
processes that cause wave dissipation and heating of the coronal plasma.

The RMHD equations can also be formulated in terms of Elsasser-like variables,
$\omega_{\pm} \equiv \omega \mp v_A \alpha$, where $\omega_{+}$ and
$\omega_{-}$ are the vorticities of the dominant and minority waves,
respectively \citep[e.g.,][]{Perez2013}. For our spectral decomposition of the
wave patterns, the wave equations are
\begin{equation}
\frac{\partial \omega_{\pm,k}} {\partial t} = - ( u_0 \pm v_A ) \frac{\partial
\omega_{\pm,k} } {\partial r} - \left[ \frac{dv_A} {dr} \pm \frac{u_0}{2 H_\rho}
\right] v_A \alpha_k + \frac{u_0} {H_{\rm B}} \omega_k + \beta D^6 \omega_{\pm,k}
+ \cdots ,  \label{eq:omeg_pm}
\end{equation}
where $\omega_{\pm,k} (r,t) \equiv \omega_k \mp v_A \alpha_k$ are the vorticity
amplitudes of the individual modes, $H_\rho (r) \equiv \rho_0 / (d \rho_0 /dr)$
is the density scale height, and the dots indicate nonlinear and $\nu_k$-damping
terms. The first term on the right-hand side of equation (\ref{eq:omeg_pm})
describes the effects of wave propagation. Note that in the
region beyond the Alfv\'{e}n critical point (where $u_0 > v_A$) the ``inward" waves
are actually carried outward by the flow. The second and third terms affect the
amplification of the waves as they propagate outward or inward, and also include
the linear couplings between the two modes. Note that these couplings
occur only between modes with the same transverse wave pattern (indicated by
index $k$). It can be shown that equation (\ref{eq:omeg_pm}) is equivalent to
equation (14) of \citet[][]{Perez2013}.

The RMHD equations are solved numerically, using the finite-difference method
for the radial derivatives. The radial grid $r_n$ has 10,892 grid points. The
grid is chosen such that the outward-wave propagation time between neighboring
grid points is constant, $( r_{n+1} - r_n ) / ( u_0 + v_A )_{n+1/2} = \Delta t$,
independent of $n$, where $\Delta t = 1$ s is the time step of the simulation.
For each time step, we first compute the change in $h_k$ and $\omega_k$ due to
wave propagation and reflection, using equation (\ref{eq:omeg_pm}). The first
term in this equation describes wave propagation, and its effect is evaluated
using the method of characteristics:
\begin{eqnarray}
\omega_{+,k} (\tilde{x}_n, t+\Delta t) & \approx &
\omega_{+,k} (\tilde{x}_n - \Delta t,t) =
\omega_{+,k} (\tilde{x}_{n-1},t) ,  \label{eq:omeg_p}  \\
\omega_{-,k} (\tilde{x}_n, t+\Delta t) & \approx &
\omega_{-,k} (\tilde{x}_n + \tilde{f}_n \Delta t,t) ,  \label{eq:omeg_m}
\end{eqnarray}
where $\tilde{x}_n$ denotes the (outward) wave travel time at position $r_n$,
and $\tilde{f}_n$ is the ratio of inward and outward wave speeds:
\begin{equation}
\tilde{f}_n \equiv \frac{v_A (r_n) - u_0 (r_n)} {v_A (r_n) + u_0 (r_n)} .
\end{equation}
Equation (\ref{eq:omeg_p}) shows that the dominant, outward-propagating waves
$\omega_{+,k}$ simply move from one grid point to the next; this allows such
waves to travel to large height in the model without any distortion of their
radial profiles. However, for the minority waves we must use interpolation,
and we use a fourth-order interpolation scheme:
\begin{equation}
\omega_{-,k} (\tilde{x}_n, t+\Delta t) \approx \omega_{-,k} (\tilde{x}_n, t)
+ a_n \tilde{f}_n + b_n \tilde{f}_n^2 + c_n \tilde{f}_n^3 + d_n \tilde{f}_n^4 ,
\label{eq:omeg_m2}
\end{equation}
where the coefficients $a_n$, $b_n$, $c_n$ and $d_n$ are determined from the
values of $\omega_{-,i}$ at grid points $i = n-2, \cdots , n+2$ (we omit the
detailed expressions). Near the inner and outer boundaries of the model we use
quadratic or linear interpolation instead. Then the effects of the other
linear terms in equation (\ref{eq:omeg_pm}) are added, and the result is
converted to $h_k$ and $\omega_k$. Finally, we compute the change in $h_k$ and
$\omega_k$ due to the nonlinear and $\nu_k$-damping terms in equations
(\ref{eq:dhdt1}) and (\ref{eq:dodt1}). This is done by integrating these
equations over the time interval $[t,t+\Delta t]$, using a fourth-order
Runge-Kutta method. The latter uses a variable time step that is often much
smaller than $\Delta t$.

\section{Energy Equations for Waves and Plasma}

We first consider the energy equation for the waves. Let $U_{\rm mag} (r,t)$
be the magnetic energy density of the waves, $| {\bf B}_1 |^2 / 8 \pi$,
averaged over the cross-section of the flux tube.
Similarly, let $U_{\rm kin} (r,t)$ be the kinetic energy density of the waves,
$\onehalf \rho_0 | {\bf v}_1 |^2$, averaged over the cross-section.
In our RMHD model, these energy densities can be written as
sums over eigenmodes:
\begin{eqnarray}
U_{\rm mag} (r,t) & = & \frac{B_0^2} {8 \pi R^2} \sum_{k=1}^N a_k^2 h_k^2 ,
\label{eq:UB} \\
U_{\rm kin} (r,t) & = & \frac{\rho_0}{2 R^2} \sum_{k=1}^N a_k^2 f_k^2 .
\label{eq:UK}
\end{eqnarray}
Multiplying equation (\ref{eq:dhdt1}) by $B_0^2 /(4 \pi)(a_k/R)^2 h_k$
and summing over $k$, we obtain an equation for the time derivative of
$U_{\rm mag}$, and multiplying (\ref{eq:dodt1}) by $\rho_0 f_k$ we obtain the
time derivative of $U_{\rm kin}$. Adding these two equations, we find that the
nonlinear terms drop out of the equation:
\begin{equation}
\frac{\partial U_{\rm A}} {\partial t} + B_0 \frac{\partial} {\partial r}
\left( \frac{ F_{\rm A} + U_{\rm A} u_0 } {B_0} \right) = 
- U_{\rm mag} \frac{d u_0} {dr} + U_{\rm kin} \frac{u_0} {H_{\rm B}}
- Q_{\rm tot} , \label{eq:UA}
\end{equation}
where $U_{\rm A} (r,t) \equiv U_{\rm mag} + U_{\rm kin}$ is the total energy
density of the waves, and $F_{\rm A} (r,t)$ is defined by
\begin{equation}
F_{\rm A} (r,t) \equiv - \frac{B_0^2} {4 \pi R^2}
\sum_{k=1}^N a_k^2 h_k f_k .  \label{eq:FA} 
\end{equation}
The total dissipation rate $Q_{\rm tot} (r,t)$ has two contributions:
\begin{equation}
Q_{\rm tot} (r,t) \equiv Q_{\perp} + Q_{\parallel} ,  \label{eq:Qtot}
\end{equation}
where
\begin{eqnarray}
Q_{\perp} (r,t) & = & \frac{\rho_0} {R^2}
\sum_{k=1}^N a_k^2 \nu_k ( f_k^2 + v_A^2 h_k^2 ) ,  \label{eq:Qperp} \\
Q_{\parallel} (r,t) & = & \beta \frac{\rho_0} {R^2}
\sum_{k=1}^N a_k^2 [ ( D^3 f_k )^2 + v_A^2 ( D^3 h_k )^2 ].  \label{eq:Qpara}
\end{eqnarray}
The terms with $\nu_k$ describe damping at high perpendicular wavenumber,
while those with $\beta$ describe damping at high parallel wavenumber.
In deriving equation (\ref{eq:UA}) we neglected the contributions of $\beta$
terms to the energy flux. Equation (\ref{eq:UA}) is valid for arbitrary
non-WKB wave propagation.

We now consider the equations for the background atmosphere. These are
obtained by averaging the MHD equations over the cross-section of the flux tube,
and over time. Then the mass conservation equation (\ref{eq:mass}) yields
\begin{equation}
\rho_0 u_0 / B_0 = \hbox{constant} , \label{eq:mass0}
\end{equation}
and the equation of motion (\ref{eq:dvdt}) yields
\begin{equation}
\rho_0 u_0 \frac{d u_0} {dr} = - \frac{d p_0} {dr} + D_{\rm wp}
- \rho_0 \frac{G M_\odot} {r^2} .  \label{eq:dudt0} 
\end{equation}
Here $p_0$ is the plasma pressure, and $D_{\rm wp} (r)$ is the wave pressure
force. For an ideal gas $p_0 = c_1 \rho_0 T_0$, and assuming a helium abundance
of $10 \%$, $c_1 = 2.3 k_{\rm B} / (1.4 m_{\rm H})$, where $k_{\rm B}$ is the
Boltzmann constant and $m_{\rm H}$ is the hydrogen mass. For non-WKB Alfv\'{e}n
waves, the wave pressure force is given by \citep[][]{Heinemann1980,
Cranmer2005}:
\begin{equation}
D_{\rm wp} (r) = - \frac{d U_{\rm mag}} {dr} 
+ \frac{U_{\rm mag} - U_{\rm kin}} {H_{\rm B}} . \label{eq:Dwp}
\end{equation}
Here $U_{\rm mag} (r)$ and $U_{\rm kin} (r)$ are the time-averaged versions
of the quantities given in equations (\ref{eq:UB}) and (\ref{eq:UK}).
The heat equation (\ref{eq:heat}) can be written as
\begin{equation}
Q_{\rm A} = Q_{\rm adv} + Q_{\rm rad} + Q_{\rm cond} ,  \label{eq:heat0}
\end{equation}
where $Q_{\rm A} (r)$ is the time-averaged heating rate; 
$Q_{\rm adv} (r)$ is the time average of the advection terms [left-hand side
of equation (\ref{eq:heat})]; and $Q_{\rm rad} (r)$ and $Q_{\rm cond} (r)$ are
the energy loss rates due to radiation and thermal conduction.
These quantities are given by
\begin{eqnarray}
Q_{\rm adv} (r) & = & c_1 \rho_0 u_0 \left( \frac{1}{\gamma-1} \frac{d T_0}{dr}
- \frac{T_0}{\rho_0} \frac{d \rho_0}{dr} \right) ,  \label{eq:Qadv} \\
Q_{\rm rad} (r) & = & n_{\rm e} n_{\rm H} \Lambda (T_0) , \label{eq:Qrad} \\
Q_{\rm cond} (r) & = & B_0 \frac{d}{dr} \left( \frac{F_{\rm cond}} {B_0}
\right) , \label{eq:Qcond}
\end{eqnarray}
where $n_{\rm H} (r) = \rho_0 / (1.4 m_{\rm H})$ is the hydrogen density,
$n_{\rm e} (r) = 1.2 n_{\rm H}$ is the electron density, and $\Lambda (T)$
is the radiative loss function \citep[taken from Figure 1 in][]{Cranmer2007}.
Following Cranmer et al., we use a ``bridging law" for the parallel component
of the thermal conductive flux:
\begin{equation}
F_{\rm cond} (r) = \frac{\nu_{\rm coll} F_{\rm SH} + \nu_{\rm exp} F_{\rm FS}}
{\nu_{\rm coll} + \nu_{\rm exp}} ,
\end{equation}
where $F_{\rm SH} (r) \equiv - \kappa dT_0 /dr$ is the classical
Spitzer-Harm prescription for thermal conduction, and 
$F_{\rm FS} (r) = 1.5 \alpha_c n_{\rm e} u_0 k_{\rm B} T_0$ 
is free-streaming heat flux
that applies in the collisionless limit (we use $\alpha_c = 4$). Also,
$\nu_{\rm coll} (r)$ is the electron-electron collision frequency, and
$\nu_{\rm exp} = u_0 / | H_\rho |$ is the wind expansion rate. Note that
the conductivity depends strongly on temperature, $\kappa \propto T_0^{5/2}$
\citep[for details, see][]{Cranmer2007}.

Multiplying equation (\ref{eq:dudt0}) by $u_0$ and adding equation
(\ref{eq:heat0}), we obtain the energy equation for the plasma:
\begin{equation}
B_0 \frac{d}{dr} \left( \frac{F_{\rm plasma}}{B_0} \right) =
Q_{\rm A} - Q_{\rm rad} + u_0 D_{\rm wp} ,  \label{eq:E1}
\end{equation}
where $F_{\rm plasma}$ is the energy flux carried by the plasma:
\begin{equation}
F_{\rm plasma} (r) = \frac{1}{2} \rho_0 u_0^3 + \frac{\gamma}{\gamma-1} p_0 u_0
- \rho_0 u_0 \frac{G M_\odot}{r} + F_{\rm cond} . \label{eq:Fplasma}
\end{equation}
The four terms on the right-hand side represent the kinetic energy flux of the
wind, the enthalpy flux, the gravitational energy flux, and the conductive flux,
respectively. The energy equation for the waves is obtained by time-averaging
equation (\ref{eq:UA}) and rearranging terms:
\begin{equation}
B_0 \frac{d} {dr} \left( \frac{F_{\rm waves}} {B_0} \right) =
- u_0 D_{\rm wp} - Q_{\rm tot} ,
\label{eq:E2}
\end{equation}
where $D_{\rm wp} (r)$ is given by equation (\ref{eq:Dwp}), and $F_{\rm waves}$
is the energy flux carried by the Alfv\'{e}n waves \citep[][]{Heinemann1980,
Cranmer2005}:
\begin{equation}
F_{\rm waves} (r) \equiv F_{\rm A} + (2 U_{\rm mag} + U_{\rm kin} ) u_0 .
\label{eq:Fwav}
\end{equation}
Here $F_{\rm A} (r)$ and $Q_{\rm tot} (r)$ are the time averages of the
quantities defined in equations (\ref{eq:FA}) and (\ref{eq:Qtot}).

\section{Setting Up the Background Atmosphere}

Three-dimensional RMHD modeling for a thin flux tube requires that we first
set up a one-dimensional model for the background atmosphere inside the tube,
i.e., a model for the magnetic field strength $B_0 (r)$, density $\rho_0 (r)$
and outflow velocity $u_0 (r)$ as functions of position along the flux tube.
We require that this model satisfy not only the mass conservation equation
(\ref{eq:mass0}) but also the equation of motion (\ref{eq:dudt0}). The latter
includes the wave pressure force $D_{\rm wp}(r)$, which plays an important role
in producing the fast solar wind emanating from coronal holes.

Following \citet[][]{Cranmer2007}, we approximate the wave pressure force by
assuming that the dominant waves are much stronger than the minority waves,
$| \omega_{+} | \gg | \omega_{-} |$. Then the mode amplitudes
for velocity and magnetic field are highly correlated, $f_k \approx - v_A h_k$,
and using this expression in equations (\ref{eq:UK}) and (\ref{eq:FA}) we find
\begin{eqnarray}
U_{\rm kin} & \approx & U_{\rm mag} \approx \onehalf U_{\rm A} , \label{eq:UK1} \\
F_{\rm A} & \approx & 2 v_A U_{\rm mag} \approx v_A U_{\rm A} . \label{eq:FA1}
\end{eqnarray}
Inserting these approximations into the time-averaged version of the wave
energy equation (\ref{eq:UA}) yields the so-called wave action equation:
\begin{equation}
\frac{d S_{\rm A}} {dr} = - ( 1 + M_A ) \frac{Q_{\rm A}} {B_0} ,
\label{eq:dSA}
\end{equation}
where $S_{\rm A} (r)$ is the wave action per unit magnetic flux within the
tube:
\begin{equation}
S_{\rm A} (r) \equiv \frac{( u_0 + v_A )^2 U_{\rm A}} {B_0 v_A} =
( 1 + M_A )^2 \frac{ U_{\rm A} } {\sqrt{4 \pi \rho_0}} ,
\label{eq:SA}
\end{equation}
and $M_A (r) \equiv u_0 / v_A$ is the Alfv\'{e}n Mach number. Here we use
$M_A \propto \rho_0^{-1/2}$, which follows from equation (\ref{eq:mass0}).
Inserting (\ref{eq:UK1}) into equation (\ref{eq:Dwp}) yields $D_{\rm wp}
\approx - \onehalf dU_{\rm A}/dr$, and the derivative in this expression
can be computed from equations (\ref{eq:SA}) and (\ref{eq:dSA}). This
yields the following expression for the wave pressure force:
\begin{equation}
D_{\rm wp} \approx - W_U \frac{d \rho_0} {dr} + \rho_0 W_Q ,
\label{eq:Dwp1}
\end{equation}
where $W_U$ and $W_Q$ are defined by
\begin{eqnarray}
W_U (r) & \equiv & \frac{U_{\rm A}} {4 \rho_0} \left( \frac{1+3 M_A}{1+M_A}
\right) , \label{eq:WU} \\
W_Q (r) & \equiv & \frac{Q_{\rm A}} {2 \rho_0 ( u_0 + v_A )} .
\label{eq:WQ}
\end{eqnarray}
Inserting expression (\ref{eq:Dwp1}) into equation (\ref{eq:dudt0}) and
using mass conservation to eliminate the density, we find the so-called
wind equation:
\begin{equation}
\left( u_0 - \frac{c_1 T_0 + W_U} {u_0} \right) \frac{du_0} {dr} =
- \frac{c_1}{B_0} \frac{d}{dr} \left( B_0 T_0 \right) 
- \frac{W_U}{B_0} \frac{dB_0} {dr} + W_Q
- \frac{G M_\odot} {r^2} ,  \label{eq:wind}
\end{equation}
consistent with equation (58) of \citet[][]{Cranmer2007}.
For the models considered in this paper, the temperature $T_0 (r)$ is
a known function of position, see equation (\ref{eq:T0}) below. Assuming
the functions $W_U (r)$ and $W_Q(r)$ are also known, we can solve equation
(\ref{eq:wind}) in a standard way: first find the position of the critical
point $r_{\rm c}$ where the right-hand side of equation (\ref{eq:wind})
vanishes; then integrate equation (\ref{eq:wind}) upward and downward in
height, starting from points just above and below the critical point,
respectively. This yields the outflow velocity $u_0 (r)$ at all heights.
The density $\rho_0 (r)$ can then be computed by using mass flux
conservation and the boundary condition on density at the coronal base.

Since the functions $W_U (r)$ and $W_Q(r)$ are not known {\it a priori},
we must determine them iteratively. In each iteration we treat these
quantities as known functions, and we solve the wind equation in the
standard way (in the first iteration we set $W_U = W_Q = 0$). This yields
new or updated values for the outflow velocity $u_0 (r)$ and density
$\rho_0 (r)$ as described above. We then compute the energy loss rates
$Q_{\rm adv} (r)$, $Q_{\rm rad} (r)$ and $Q_{\rm cond} (r)$, and using
equation (\ref{eq:heat0}) we obtain an improved estimate for the heating
rate $Q_{\rm A} (r)$ needed to sustain the background atmosphere.
Next, we integrate equation (\ref{eq:dSA}) from the base upward.
This yields the wave action parameter $S_{\rm A} (r)$, from which we can
determine the wave energy density $U_{\rm A} (r)$. Finally, we recompute
$W_U (r)$ and $W_Q (r)$ from equations (\ref{eq:WU}) and (\ref{eq:WQ}),
and we repeat the iterative process, until the changes in $W_U$ and $W_Q$
become sufficiently small.

The temperature $T_0 (r)$ must be specified in such a way that a critical
point can always be found. For a proper critical point to exist, the
function $F(r)$ on the right-hand side of equation (\ref{eq:wind}) must
have a root, and the slope of the function at the root must be positive,
$(dF/dr)_{\rm c} > 0$. In particular, the first term in $F(r)$ related to
temperature must be positive, and must decrease with $r$ at a rate which
is less than that of the gravity term, $G M_\odot / r^2$. To ensure that
this condition is satisfied, we specify not the temperature itself but
rather the first term in $F(r)$:
\begin{equation}
- \frac{c_1}{B_0} \frac{d}{dr} \left( B_0 T_0 \right) =
C_0 \frac{G M_\odot}{R_\odot^2} \left( \frac{r}{R_\odot} \right)^{-m-1}
\left[ 1 - C_1 \left( \frac{r}{R_\odot} \right)^{-k} \right] ,
\label{eq:dBT}
\end{equation}
where $C_0$ and $C_1$ are dimensionless constants. For the first term
on the right-hand side of equation (\ref{eq:dBT}) to decreases more
slowly than the gravity term in the wind equation, we require that the
exponent $m < 1$. Inserting equation (\ref{eq:B0poten}) into
(\ref{eq:dBT}), we find for the temperature
\begin{equation}
T_0 (r) = \frac{G M_\odot}{c_1 R_\odot} \frac{C_0} {B_0 (r)} \sum_{n=1}^5
B_n \left[ \frac{1}{2n+m} \left( \frac{r}{R_\odot} \right)^{-2n-m} -
\frac{C_1}{2n+m+k} \left( \frac{r}{R_\odot} \right)^{-2n-m-k} \right] .
\label{eq:T0}
\end{equation}
For the temperature to decrease with $r$ at large height, we require
$m > 0$. In the present paper we use $C_0 = 0.35$, $C_1 = 2$, $m = 0.3$
and $k = 8$.

\clearpage

\end{document}